\documentclass{article}

\usepackage{arxiv}

\usepackage[utf8]{inputenc} % allow utf-8 input
\usepackage[T1]{fontenc}    % use 8-bit T1 fonts
\usepackage{hyperref}       % hyperlinks
\usepackage{url}            % simple URL typesetting
\usepackage{doi}
\usepackage[table,xcdraw]{xcolor}
\usepackage{amsmath,amsfonts}
\usepackage{algorithmic}
\usepackage{balance}
\usepackage{graphicx}
\usepackage{textcomp}
\usepackage{todonotes}
\usepackage{hyperref}
\usepackage{float}
\usepackage{comment}

\usepackage{algorithm}
\usepackage{lscape}
\usepackage{longtable}
\usepackage{subcaption}
\usepackage{multirow}

\usepackage[many]{tcolorbox} 
\usepackage{tabularx}
\usepackage{booktabs}

\usepackage[backend=biber,natbib,mincitenames=1, maxcitenames=1, uniquelist=true,style=acmnumeric,giveninits=true]{biblatex}
\newcommand{\rqone}{Does temporal proximity impact the accuracy of TLP?}
\newcommand{\rqtwo}{Does temporal proximity impact the power of TLP features?}

\title{Anticipating Bugs: Ticket-Level Bug Prediction and Temporal Proximity Effects}

\date{}

\newif\ifuniqueAffiliation
\uniqueAffiliationtrue

\ifuniqueAffiliation % Standard variant of author block
\author{ 
	\href{https://orcid.org/0009-0000-9308-2508}{\includegraphics[scale=0.06]{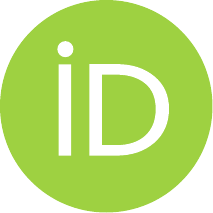}\hspace{1mm}Daniele La Prova} \\
	University of Rome "Tor Vergata"\\
	Via del Politecnico, 1, Rome, Italy, 00133  \\
	\texttt{daniele.laprova@hotmail.it} \\
	\And
	\href{https://orcid.org/0009-0002-4283-9114}{\includegraphics[scale=0.06]{orcid.pdf}\hspace{1mm}Emanuele Gentili} \\
	University of Rome "Tor Vergata"\\
	Via del Politecnico, 1, Rome, Italy, 00133  \\
	\& \\
	MBDA Italy Spa\\
	Via Monte Flavio,  45, Rome, Italy, 00131  \\
	\texttt{emanuele.gentili@mbda.it} \\
	\And
	\href{https://orcid.org/0000-0002-6340-0058}{\includegraphics[scale=0.06]{orcid.pdf}\hspace{1mm}Davide Falessi} \\
	University of Rome "Tor Vergata"\\
	Via del Politecnico, 1, Rome, Italy, 00133  \\
	\texttt{falessi@ing.uniroma2.it} \\
}
\else
\usepackage{authblk}

\newbox{\orcid}\sbox{\orcid}{\includegraphics[scale=0.06]{orcid.pdf}} 
\author[1]{%
	\href{https://orcid.org/0009-0000-9308-2508}{\includegraphics[scale=0.06]{orcid.pdf}\hspace{1mm}Daniele La Prova} %\\
}
\author[1,2]{%
	\href{https://orcid.org/0009-0002-4283-9114}{\includegraphics[scale=0.06]{orcid.pdf}\hspace{1mm}Emanuele Gentili} %\\
}
\author[1]{%
	\href{https://orcid.org/0000-0002-6340-0058}{\includegraphics[scale=0.06]{orcid.pdf}\hspace{1mm}Davide Falessi}%
}

\affil[1]{University of Rome "Tor Vergata", Via del Politecnico, 1, Rome, Italy, 00133}
\affil[2]{MBDA Italy Spa, Via Monte Flavio,  45, Rome, Italy, 00131}
\fi

\renewcommand{\shorttitle}{Anticipating Bugs: Ticket-Level Bug Prediction and Temporal Proximity Effects}

\hypersetup{
pdftitle={Anticipating Bugs: Ticket-Level Bug Prediction and Temporal Proximity Effects},
pdfsubject={q-bio.NC, q-bio.QM},
pdfauthor={Daniele La Prova, Emanuele Gentili, Davide Falessi},
pdfkeywords={Defects, Defect Prediction, Machine Learning for Software Engineering},
}

\begin{document}
\maketitle

\begin{abstract}
\textbf{Context}: Software bugs significantly impact project time, budgets, and safety, motivating extensive research in bug prediction. The primary goal of bug prediction is to optimize testing efforts by focusing on software fragments, i.e., classes, methods, commits (JIT), or lines of code, most likely to be buggy. However, these predicted fragments already contain bugs. Thus, the current bug prediction approaches support fixing rather than prevention.

\textbf{Aim}: Motivated by the principle of "prevention is better than cure," the aim of this paper is to introduce and evaluate Ticket-Level Prediction (TLP), an approach to identify tickets that will introduce bugs once implemented. We analyze TLP at three temporal points, each point represents a ticket lifecycle stage: Open, In Progress, or Closed. We conjecture that: (1) TLP accuracy increases as tickets progress towards the closed stage due to improved feature reliability over time, and (2) the predictive power of features changes across these temporal points.  Our TLP approach leverages 72 features belonging to six different families: code, developer, external temperature, internal temperature, intrinsic, ticket to tickets, and JIT.

\textbf{Method}: Our TLP evaluation uses a sliding-window approach, balancing feature selection and three machine-learning bug prediction classifiers on about 10,000 tickets of two Apache open-source projects. 
  
\textbf{Results}: Our results show that TLP accuracy increases with proximity, confirming the expected trade-off between early prediction and accuracy. Regarding the prediction power of feature families, no single feature family dominates across stages; developer-centric signals are most informative early, whereas code and JIT metrics prevail near closure, and temperature-based features provide complementary value throughout.

\textbf{Conclusions}: Our findings complement and extend the literature on bug prediction at the class, method, or commit level by showing that defect prediction can be effectively moved upstream, offering opportunities for risk-aware ticket triaging and developer assignment before any code is written.
  
\end{abstract}

% keywords can be removed

\section{Introduction}
Software engineering comes into play here as the discipline that deals with the cost efficiency, good quality and timely delivery of software systems using quantitative and measurable approaches \citep{ghezzi1991fundamentals}. When developing a project using SE principles, it is not uncommon that developers organize requirements in tickets since they are a good way to keep track of the work \citep{smith2016managing}.

Since bugs can have a significant impact on time, budget and safety, much effort has been spent on bug prediction. The main purpose of bug prediction is to minimize the testing effort. This is achieved by focusing testing on specific software artifacts, such as classes, methods, commits, or lines of code predicted to be buggy. Consequently, significant progress has been made in developing prediction models at the class, method, commit, and line levels \cite{DBLP:conf/wcre/KameiS16, DBLP:journals/csur/ZhaoDC23, DBLP:journals/csur/ZhaoDC23,DBLP:journals/ase/LiDZJW24, DBLP:journals/tse/Tantithamthavorn20, DBLP:journals/infsof/FuMS16, DBLP:journals/tse/Tantithamthavorn19, DBLP:journals/tosem/FalessiAP22, DBLP:journals/tse/McIntoshK18, DBLP:journals/tse/SongM23, DBLP:journals/ese/FalessiLCEC23,DBLP:journals/jss/OzakinciT18}. However, these predicted entities already contain bugs.

Change impact analysis is a software engineering research area focusing on  assessing the consequences of modifications to software systems so that teams can minimize unintended consequences, optimize their development efforts, and establish maintenance and testing strategies \cite{ DBLP:conf/iwpse/Lehnert11, DBLP:conf/sbes/BordinB18,  DBLP:journals/access/AnwerWWM19, DBLP:conf/icsoft/GentiliCF24,DBLP:conf/iwpc/AungHS20}.  

Requirements quality refers to the degree to which software requirements are well-defined, unambiguous, complete, consistent, and testable \cite{ berry1998requirements, DBLP:conf/icse/WilsonRH97}.  High-quality requirements are essential for guiding development teams and ensuring the final product aligns with stakeholders' needs \cite{wiegers2013software}.  \citet{berry1998requirements} emphasize that clear and precise requirements minimize misunderstandings during development, leading to more efficient project execution, and several studies \cite{DBLP:journals/jss/AhonenS10,DBLP:books/daglib/0025717,boehm2007software} report on the disruptive effects that ambiguity, inconsistency, incompleteness or, more generally, “requirements smells” \cite{DBLP:conf/profes/GentiliF23} can have on SW project success \cite{DBLP:conf/re/KamataT07}.  

Models prediction accuracy is impacted by time \cite{box2015time,DBLP:conf/ijcnn/TaiebBSL09,brockwell2002introduction,orrell2001model}, and it is reasonable to assume that the closer we get to the prediction instant, the more information we gain, and the more precise the prediction becomes. So, the concept of temporal proximity in prediction plays a significative role, as prediction accuracy typically declines over longer time horizons due to the error accumulation of long-term predictors \cite{ box2015time, orrell2001model}, especially when applied to intrinsically stochastic processes, like weather forecasting \cite{brockwell2002introduction,lorenz1963deterministic}, financial stock market \cite{lo2011non}, or epidemic modelling \cite{keeling2008modeling}.

With the idea that prevention is better than cure, our aim is to propose and evaluate a first approach for ticket-level prediction (TLP); the approach predicts which tickets, when implemented, will lead to bug injection. 

We consider three temporal points to characterize the lifecycle of a ticket: created, assigned, and implemented. We investigate how temporal proximity impacts TLP in terms of the accuracy and power of the predictive features. Specifically, we conjecture that: 1) TLP accuracy improves as the ticket moves closer to implementation (i.e., the bug moves closer to its injection) due to the increasing reliability of predictive features over time, and 2) the power of predictive features changes over time.

As TLP features, we propose and measure 72 features coming from commit-level and class-level defect prediction, requirements quality, NLP, and the broader software engineering domain.

Our TLP evaluation considers balancing, feature selection, and many machine-learning bug prediction classifiers on about 11,000 tickets related to two open-source projects from the Apache ecosystem. As TLP accuracy metrics we use Precision, Recall, F1, AUC, Kappa and GMean. As TLP power metrics we use the info gain ratio and backward search feature selection.

Our results show that TLP accuracy increases with proximity, confirming the expected trade-off between early prediction and accuracy. Regarding the prediction power of feature families, no single feature family dominates across stages; developer-centric signals are most informative early, whereas code and JIT metrics prevail near closure, and temperature-based features provide complementary value throughout.

Our findings complement and extend the literature on bug prediction at the class, method, or commit level by showing that defect prediction can be effectively moved upstream, offering opportunities for risk-aware ticket triaging and developer assignment before any code is written.

Thus, the contribution of this paper is threefold:
\begin{enumerate}
\item We propose a novel framework for predicting defect-prone tickets before any code is implemented.
\item We empirically evaluate the framework across key stages of the ticket lifecycle, highlighting the trade-off between early prediction and accuracy.
\item We conduct a comparative analysis of 72 features spanning defect prediction metrics, natural language processing (NLP), and requirements quality.
\end{enumerate}

The remainder of this paper is structured as it follows. Section \ref{sec:related} describes how this paper position with past works. Section \ref{sec:design} describes the empirical study design. Section \ref{sec:results} reports the results and Section \ref{sec:discussions} discusses them. Section \ref{sec:threats} discusses the threats to the study validity.  Finally, section \ref{sec:conclusion} concludes the paper and outlines directions for future work.

\section{Ticket-level Prediction Features}\label{sec:features}
This section presents the features as organized into coherent families, each motivated by prior defect prediction literature across multiple granularities (e.g., class-, method-, line-, and ticket-level) \cite{DBLP:journals/ese/FalessiLCEC23, DBLP:journals/ase/LiDZJW24, DBLP:journals/jss/OzakinciT18}. We adopted a snowballing strategy \cite{DBLP:conf/ease/Wohlin14} to expand the initial set of sources and further incorporated findings from requirements engineering research to identify text- and ticket-based indicators. The final feature set comprises 72 features, grouped into seven semantically coherent families. In this section, we describe the rationale behind each family and provide a few examples. Moreover, for each family we report a summary table describing for each feature the name, implementation code identifier, bibliographic reference, and measurement timing; note that feature values may change over the lifecycle of the ticket.
Appendix \ref{sec:Featurefamilies} provides a complete list and definition of each feature.

\subsection{Code}
These features aim to capture the intrinsic characteristics of the underlying system that influence the ease and risk of implementing a ticket, see \autoref{tab:tlp_ff_code} \cite{DBLP:conf/icsm/Zhang09, DBLP:conf/wcre/KochharWL16}. This includes static quality metrics and structural attributes that correlate with maintainability and defect-proneness. As an example of feature from this family, we assess the internal quality of the code by counting the number of rule violations detected by PMD \cite{copeland2005pmd, DBLP:conf/esem/FalessiRM17}, which serves as a proxy for code smells across categories such as design, documentation, error handling, multithreading, and performance \cite{DBLP:books/daglib/0019908}.
\begin{table}[h]
    \centering
    \caption{TLP features for Code (C) family.}
    \label{tab:tlp_ff_code}    
    %\begin{tabular}{@{}llll@{}}
    \small
    \begin{tabularx}{\linewidth}{XXll}
        \toprule
        \textbf{Name}                  & \textbf{CodeName}                      & \textbf{Reference}                                                                 & \textbf{Availability} \\ 
        \midrule
        Number of smells               & \textit{code\_quality-smells\_count}            & \citep{DBLP:journals/information/CairoCM18} \citep{DBLP:journals/tse/FalessiRGC20} & Open                  \\
        Number of different languages  & \textit{code\_size-number\_of\_languages}       & \citep{DBLP:conf/wcre/KochharWL16}                                                 & Open                  \\
        Number of files                & \textit{code\_size-number\_of\_files}           & \citep{DBLP:conf/icsm/Zhang09}                                                     & Open                  \\
        Total LOCs                     & \textit{code\_size-total\_LOCs}                 & \citep{DBLP:conf/icsm/Zhang09}                                                     & Open                  \\
        \bottomrule
    \end{tabularx}
\end{table}

\subsection{Developer}
These features aim to capture the historical performance and familiarity of the ticket assignee, in line with prior work linking developer experience to code quality see \autoref{tab:tlp_ff_developer} \cite{DBLP:journals/jss/WangLCHZX20,DBLP:conf/promise/MatsumotoKMMN10}. For instance, we consider the number of tickets assigned to the developer in the past, which serves as a proxy for their experience and familiarity with the codebase. This feature is based on the assumption that developers with more experience are less likely to introduce defects \cite{DBLP:journals/ese/PatelAH24}.
\begin{table}[h]
    \centering
    \caption{TLP features for Developer (D) family.}
    \label{tab:tlp_ff_developer}
    %\begin{tabular}{@{}llll@{}}
    \small
    \begin{tabularx}{\linewidth}{XXll}
        \toprule
        \textbf{Name}                     & \textbf{CodeName}         & \textbf{Reference}                        & \textbf{Availability} \\ 
        \midrule
        Assigned Developer's ANFIC        & \textit{assignee-ANFIC}            & \citep{DBLP:conf/promise/MatsumotoKMMN10} & Assigned              \\
        Assigned Developer's Familiarity  & \textit{assignee-familiarity}      & \citep{DBLP:journals/jss/WangLCHZX20}     & Assigned              \\
        \bottomrule
    \end{tabularx}
\end{table}

\subsection{External Temperature}
These features aim to capture the project-wide activity that may interfere with development quality , see \autoref{tab:tlp_ff_external_temperature} \cite{DBLP:journals/tosem/PerrySV01, DBLP:journals/tse/FalessiRGC20}. For instance, we consider the number of open tickets in the project, which serves as a proxy for the workload and potential distractions faced by developers. This feature is based on the assumption that a higher number of open tickets may lead to increased pressure and reduced attention to detail, potentially resulting in more defects \cite{DBLP:journals/ese/FalessiLCEC23}.
\begin{table}[h]
    \centering
    \caption{TLP features for External Temperature (E\_T) family.}
    \label{tab:tlp_ff_external_temperature}
    %\begin{tabular}{@{}llll@{}}
	\small
	\begin{tabularx}{\linewidth}{XXll}
	\toprule
    \textbf{Name}                          & \textbf{CodeName}                   & \textbf{Reference}                                                      & \textbf{Availability} \\ 
    \midrule
	Temporal Locality                      & \textit{temporal\_locality}                 & \citep{DBLP:journals/tse/FalessiRGC20} \citep{DBLP:conf/ecoop/GuH0021}  & Open     \\
	Weighted Temporal Locality             & \textit{temporal\_locality-weighted}         & \citep{DBLP:journals/tse/FalessiRGC20} \citep{DBLP:conf/ecoop/GuH0021}  & Open     \\
	Number of Commits While In Progress    & \textit{commits\_while\_in\_progress-count}  & \citep{DBLP:journals/tosem/PerrySV01}                                   & Assigned \\
	Churn of Commits While In Progress     & \textit{commits\_while\_in\_progress-churn}  & \citep{DBLP:journals/tosem/PerrySV01} \citep{DBLP:conf/scam/FaragoHF15} & Assigned \\
	Latest Commit Churn                    & \textit{latest\_commit-churn}                & \citep{DBLP:conf/scam/FaragoHF15}                                       & Assigned \\
	Latest Commit Number of Files          & \textit{latest\_commit-number\_of\_files}    & \citep{DBLP:conf/msr/KeshavarzN22}                                      & Assigned \\
	\bottomrule
    \end{tabularx}
\end{table}

\subsection{Internal Temperature}
Some file level bug-predicting approaches assume that files recently or frequently changed are more bug-prone , see \autoref{tab:tlp_ff_internal_temperature} \citep{5463279}. We transfer this concept to the ticket level by measuring the "hotness" of the ticket, namely how frequently the ticket was subject to activities and by how many different stakeholders. For instance, we consider the number of comments added to the ticket, which serves as a proxy for the level of discussion and collaboration surrounding the ticket. This feature is based on the assumption that tickets with more comments may indicate a higher level of complexity or uncertainty, potentially leading to more defects \cite{DBLP:journals/ese/FalessiLCEC23}.This family also incorporates sentiment analysis of requirement descriptions and comments, emphasizing sentence-level polarity and subjectivity as potential defect indicators \citep{DBLP:reference/ml/0016017, incose2023incose}. Building on prior findings linking sentiment to software quality and resolution speed \citep{DBLP:conf/fase/BacchelliDL10, 9307811}, we incorporate features such as the occurrence and proportion of negative comments to capture affective signals associated with defect-prone tickets.

\begin{table}[h]
    \centering
    \caption{TLP features for Internal Temperature (I\_T) family.}
    \label{tab:tlp_ff_internal_temperature}
    %\begin{tabular}{@{}llll@{}}
    \small
    \begin{tabularx}{\linewidth}{XXll}
	\toprule
    \textbf{Name}                     & \textbf{CodeName}              & \textbf{Reference}                                                     & \textbf{Availability} \\ 
    \midrule
	Ticket Participants Count          & \textit{issue\_participants-count}      & \citep{DBLP:conf/sigsoft/PinzgerNM08}                                  & Open \\
    Activities Count                   & \textit{activities-count}               & \citep{DBLP:conf/fase/BacchelliDL10}                                   & Open \\
    Comments Count                     & \textit{activities-comments\_count}     & \citep{DBLP:conf/fase/BacchelliDL10}                                   & Open \\
    Work Items Count                   & \textit{activities-work\_items\_count}  & \citep{DBLP:conf/fase/BacchelliDL10}                                   & Open \\
    Histories Count                    & \textit{activities-histories\_count}    & \citep{DBLP:conf/fase/BacchelliDL10}                                   & Open \\
    Sentiment Polarity                 & \textit{nlp4re\_sentiment-IT\_POL}      & \citep{DBLP:conf/fase/BacchelliDL10} \citep{DBLP:reference/ml/0016017} & Open \\
    Sentiment Subjectivity             & \textit{nlp4re\_sentiment-IT\_SUB}      & \citep{DBLP:conf/fase/BacchelliDL10} \citep{DBLP:reference/ml/0016017} & Open \\
    Number Of Negative Sentiment       & \textit{nlp4re\_sentiment-CM\_NNS}      & \citep{9307811}                                                        & Open \\
    Percentage Of Negative Sentiment   & \textit{nlp4re\_sentiment-CM\_PNS}      & \citep{9307811}                                                        & Open \\
    Presence Of One Negative Sentiment & \textit{nlp4re\_sentiment-CM\_ONS}      & \citep{9307811}                                                        & Open \\
	\bottomrule
    \end{tabularx}
\end{table}

\subsection{Intrinsic}

Intuitively, some tickets can be considered inherently more difficult to implement than others. These features aim to capture the intrinsic complexity of the ticket, which is not directly related to the code or the developer but rather to the nature of the ticket itself , see \autoref{tab:tlp_ff_intrinsic}. For instance, we consider the number of requirements in the ticket, which serves as a proxy for the complexity and scope of the ticket. This feature is based on the assumption that tickets with more requirements may be more complex and therefore more likely to introduce defects \cite{DBLP:journals/ese/FalessiLCEC23, DBLP:journals/ese/PatelAH24, DBLP:journals/tse/WinterBCHHNW23, DBLP:conf/icse/WilsonRH97,DBLP:conf/issre/JiangCM07}. Other features we took into account are the type of the ticket (e.g., bug, feature request, task) and the priority assigned to the ticket. These features are based on the assumption that different types of tickets may have different defect rates, and that higher-priority tickets may be more likely to introduce defects due to increased pressure and reduced attention to detail \cite{DBLP:journals/ese/FalessiLCEC23, DBLP:journals/ese/PatelAH24}.
\begin{table}[h]
    \centering
    \caption{TLP features for Intrinsic (I) family.}
    \label{tab:tlp_ff_intrinsic}
    %\begin{tabular}{@{}llll@{}}
    \small
    \begin{tabularx}{\linewidth}{XXll}
	\toprule
    \textbf{Name}                     & \textbf{CodeName}              & \textbf{Reference}                                                     & \textbf{Availability} \\ 
    \midrule
	Priority                             & \textit{priority}                      &                                                                                       & Open \\
    Component Count                      & \textit{components-count}              & \citep{DBLP:journals/ese/PatelAH24}                                                   & Open \\
    Components Max Bugginess             & \textit{components-max\_bugginess}     & \citep{DBLP:journals/ese/PatelAH24}                                                   & Open \\
    Type                                 & \textit{type}                          & \citep{DBLP:conf/icse/GuBHS10} \citep{DBLP:journals/tse/WinterBCHHNW23}               & Open \\
    Description Attribute Actions        & \textit{nlp4re\_description-DA\_ACT}   & \citep{incose2023incose}                                                              & Open \\
    Description Attribute Conditionals   & \textit{nlp4re\_description-DA\_CND}   & \citep{DBLP:conf/icse/WilsonRH97, DBLP:conf/issre/JiangCM07}                          & Open \\
    Description Attribute Continuances   & \textit{nlp4re\_description-DA\_CNT}   & \citep{DBLP:conf/icse/WilsonRH97, DBLP:conf/issre/JiangCM07}                          & Open \\
    Description Attribute Imperatives    & \textit{nlp4re\_description-DA\_IMP}   & \citep{DBLP:conf/icse/WilsonRH97, DBLP:conf/issre/JiangCM07}                          & Open \\
    Description Attribute incompletes    & \textit{nlp4re\_description-DA\_INC}   & \citep{DBLP:conf/icse/WilsonRH97, DBLP:conf/issre/JiangCM07}                          & Open \\
    Description Attribute Options        & \textit{nlp4re\_description-DA\_OPT}   & \citep{DBLP:conf/icse/WilsonRH97, DBLP:conf/issre/JiangCM07}                          & Open \\
    Description Attribute Sources        & \textit{nlp4re\_description-DA\_SRC}   & \citep{incose2023incose}                                                              & Open \\
    Description Attribute Weak Phrases   & \textit{nlp4re\_description-DA\_WKP}   & \citep{DBLP:conf/icse/WilsonRH97, DBLP:conf/issre/JiangCM07}                          & Open \\
    Description Attribute Risk Level     & \textit{nlp4re\_description-DA\_RKL}   & \citep{incose2023incose} \citep{DBLP:conf/icse/WilsonRH97, DBLP:conf/issre/JiangCM07} & Open \\
    Number Of subjects                   & \textit{nlp4re\_description-EX\_SBJ}   & \citep{incose2023incose}                                                              & Open \\
    Number Of words                      & \textit{nlp4re\_description-EX\_CNS}   & \citep{DBLP:conf/icse/WilsonRH97}                                                     & Open \\
    Number Of Verbs                      & \textit{nlp4re\_description-EX\_VRB}   & \citep{incose2023incose}                                                              & Open \\
    Number Of Ambiguities                & \textit{nlp4re\_description-EX\_AMG}   & \citep{incose2023incose}                                                              & Open \\
    Number Of Directives                 & \textit{nlp4re\_description-EX\_DIR}   & \citep{DBLP:conf/icse/WilsonRH97}                                                     & Open \\
    Readability Score                    & \textit{nlp4re\_description-EX\_RDS}   & \citep{DBLP:conf/icse/WilsonRH97}                                                     & Open \\
    Sentences Completeness               & \textit{nlp4re\_description-EX\_ICP}   & \citep{DBLP:conf/icse/WilsonRH97}                                                     & Open \\
    Action Density                       & \textit{nlp4re\_description-EX\_ACD}   & \citep{incose2023incose}                                                              & Open \\
    Number Of Entities                   & \textit{nlp4re\_description-EX\_ENT}   & \citep{incose2023incose}                                                              & Open \\
	\bottomrule
    %\end{tabular}
    \end{tabularx}
\end{table}

\subsection{Ticket-to-Tickets Similarity (T2T)}
These features aim to capture the semantic similarity to previously bug-inducing , see \autoref{tab:tlp_ff_t2t}. The intuition is that tickets semantically similar to previously bug-inducing ones are more likely to introduce defects. Similarity is computed using three established NLP techniques: cosine similarity on TF-IDF vectors \citep{salton1988term}, Jaccard similarity on token sets \citep{manning2008introduction}, and Euclidean distance on term frequency vectors \citep{mikolov2013linguistic}. Each metric is applied to both the ticket title and description, and aggregated via maximum and average operators, leading to 12 distinct features, e.g., 3 techniques * 2 text fields * 2 aggregators.

\begin{table}[h]
    \centering
    \caption{TLP features for Ticket to Ticket (T2T) family.}
    \label{tab:tlp_ff_t2t}
    \small
    \begin{tabularx}{\linewidth}{Xlll}
	\toprule
    \textbf{Name}             & \textbf{CodeName}                                                      & \textbf{Reference}  & \textbf{Availability} \\ 
    \midrule                                                                                                                                                               
	% Average Bag of Words - Description & \textit{buggy\_similarity-avg\_similarity\_bag\_of\_words\_cosine\_text} & \citep{harris1954distributional} & Open  \\
    % Average TF-IDF - Title             & \textit{buggy\_similarity-avg\_similarity\_TFIDFcosine\_title}           & \citep{salton1988term}           & Open  \\
    % Max Levenshtein - Title            & \textit{buggy\_similarity-max\_similarity\_levenshtein\_title}           & \citep{levenshtein1966binary}    & Open  \\
    Max Jaccard Title       & \textit{buggy\_similarity-max\_similarity\_jaccard\_title}              & \citep{jaccard1901etude}  & Open \\      
    Max Jaccard Text        & \textit{buggy\_similarity-max\_similarity\_jaccard\_text}               & \citep{jaccard1901etude}  & Open \\
    Max TFIDF Title         & \textit{buggy\_similarity-max\_similarity\_tfidf\_cosine\_title}        & \citep{baeza1999modern}   & Open \\
    Max TFIDF Text          & \textit{buggy\_similarity-max\_similarity\_tfidf\_cosine\_text}         & \citep{baeza1999modern}   & Open \\
    Max Euclidean Title     & \textit{buggy\_similarity-max\_similarity\_euclidean\_distance\_title}  & \citep{baeza1999modern}   & Open \\
    Max Euclidean Text      & \textit{buggy\_similarity-max\_similarity\_euclidean\_distance\_text}   & \citep{baeza1999modern}   & Open \\
    Average Jaccard Title   & \textit{buggy\_similarity-avg\_similarity\_jaccard\_title}              & \citep{jaccard1901etude}  & Open \\
    Average Jaccard Text    & \textit{buggy\_similarity-avg\_similarity\_jaccard\_text}               & \citep{jaccard1901etude}  & Open \\
    Average TF-IDF Title    & \textit{buggy\_similarity-avg\_similarity\_tfidf\_cosine\_title}        & \citep{baeza1999modern}   & Open \\
    Average TF-IDF Text     & \textit{buggy\_similarity-avg\_similarity\_tfidf\_cosine\_title}        & \citep{baeza1999modern}   & Open \\
    Average Euclidean Title & \textit{buggy\_similarity-avg\_similarity\_euclidean\_distance\_title}  & \citep{baeza1999modern}   & Open \\ 
    Average Euclidean Text  & \textit{buggy\_similarity-avg\_similarity\_euclidean\_distance\_text}   & \citep{baeza1999modern}   & Open \\
	\bottomrule
    \end{tabularx}
\end{table}

\subsection{JIT}
These features are the JIT features described by \citet{DBLP:journals/tse/KameiSAHMSU13} and \citet{DBLP:conf/msr/KeshavarzN22}, see \autoref{tab:jit-features}. We neglected the feature "Year" since it is already available as a feature Author date. Since a Ticket can be linked to several commits, we aggregated aggregated using feature-specific strategies (e.g., mean, max, or sum), according to empirical behavior of the metric. 

% \usepackage{booktabs}
% \usepackage{graphicx}
\begin{table}[h]
    \centering
    \caption{JIT features.}
    \label{tab:jit-features}
    % \resizebox{\columnwidth}{!}{%
    % \begin{tabular}{@{}llll@{}}
    \small
    \begin{tabularx}{\linewidth}{XXlll}
    \toprule
    \textbf{Name}                  & \textbf{CodeName}                   & \textbf{References}                                                        & \textbf{Availability} & \textbf{Aggregation}  \\ 
    \midrule
    Authors Count                  & \textit{jit-ndev-MAX}               & \citep{DBLP:journals/tse/KameiSAHMSU13} \citep{DBLP:conf/msr/KeshavarzN22} & Closed & MAX                   \\              
    Developer Recent Experience    & \textit{jit-arexp-MIN}              & \citep{DBLP:journals/tse/KameiSAHMSU13} \citep{DBLP:conf/msr/KeshavarzN22} & Closed & MIN                   \\
    Developer Experience           & \textit{jit-aexp-MIN}               & \citep{DBLP:journals/tse/KameiSAHMSU13} \citep{DBLP:conf/msr/KeshavarzN22} & Closed & MIN                   \\
    Developer Subsystem Experience & \textit{jit-asexp-MIN}              & \citep{DBLP:journals/tse/KameiSAHMSU13} \citep{DBLP:conf/msr/KeshavarzN22} & Closed & MIN                   \\
    Modified Subsystems Count      & \textit{jit-ns-MAX}                 & \citep{DBLP:journals/tse/KameiSAHMSU13} \citep{DBLP:conf/msr/KeshavarzN22} & Closed & MAX                   \\
    Age                            & \textit{jit-age-MIN}                & \citep{DBLP:journals/tse/KameiSAHMSU13} \citep{DBLP:conf/msr/KeshavarzN22} & Closed & MIN                   \\
    Author Date                    & \textit{jit-author\_date-DURATION}  & \citep{DBLP:journals/tse/KameiSAHMSU13} \citep{DBLP:conf/msr/KeshavarzN22} & Closed & MAX(date) - MIN(date) \\
    LOCs Added                     & \textit{jit-la-SUM}                 & \citep{DBLP:journals/tse/KameiSAHMSU13} \citep{DBLP:conf/msr/KeshavarzN22} & Closed & SUM                   \\
    LOCs Deleted                   & \textit{jit-ld-SUM}                 & \citep{DBLP:journals/tse/KameiSAHMSU13} \citep{DBLP:conf/msr/KeshavarzN22} & Closed & SUM                   \\
    Type                           & \textit{jit-fix-COUNT\_TRUE}        & \citep{DBLP:journals/tse/KameiSAHMSU13} \citep{DBLP:conf/msr/KeshavarzN22} & Closed & COUNT(True)           \\
    Modified Directories Count     & \textit{jit-nd-MAX}                 & \citep{DBLP:journals/tse/KameiSAHMSU13} \citep{DBLP:conf/msr/KeshavarzN22} & Closed & MAX                   \\
    Unique Changes Count           & \textit{jit-nuc-MAX}                & \citep{DBLP:journals/tse/KameiSAHMSU13} \citep{DBLP:conf/msr/KeshavarzN22} & Closed & MAX                   \\
    Entropy                        & \textit{jit-ent-MAX}                & \citep{DBLP:journals/tse/KameiSAHMSU13} \citep{DBLP:conf/msr/KeshavarzN22} & Closed & MAX                   \\
    Modified Files Count           & \textit{jit-nf-MAX}                 & \citep{DBLP:journals/tse/KameiSAHMSU13} \citep{DBLP:conf/msr/KeshavarzN22} & Closed & MAX                   \\
    Number of Commits              & \textit{num\_commits}               & \citep{DBLP:journals/tosem/FalessiAP22}                                    & Closed & COUNT                 \\ 
    % & jit-lt                & line of code in a file before the change                          & \citep{DBLP:journals/tse/KameiSAHMSU13} \citep{DBLP:conf/msr/KeshavarzN22} & SUM                   \\
    \bottomrule
    \end{tabularx}
    %}
\end{table}

\section{Methodology}\label{sec:design}
% scelte progettuali, terminologie, alternative e considerazioni fatte

%%%% RQ1
\subsection{RQ1: \rqone}
\label{sec:design_rq1}

\subsubsection{Introduction}
\label{sec:design_rq1_intro}
In general, predicting future outcomes holds practical value, particularly when forecasts allow timely intervention. The earlier we can make an accurate prediction, the more it is actionable. Predicting a bug when already in the code limits the usefulness of the prediction. Ideally, we would like a perfect, accurate prediction as soon as a ticket is created. However, measuring features in the early stages of the ticket lifecycle could yield no or inaccurate data. Thus, in this RQ we are interested in exploring the tradeoffs between the temporal proximity of the prediction and its accuracy. Specifically, we conjecture that TLP accuracy increases as tickets progress towards the closed stage due to improved feature reliability.

\subsubsection{Independent Variables}
\label{sec:design_rq1_iv}
\begin{figure}
    \centering
    \includegraphics[width=\linewidth]{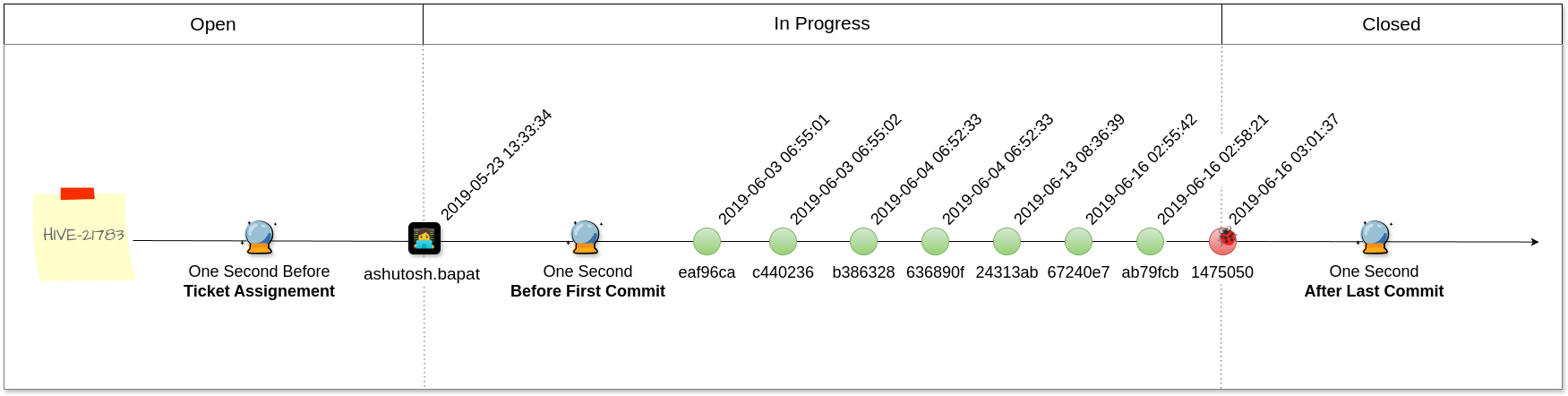}
    \caption{Example of the proximity points of the HIVE-21783 ticket, its commits and lifecycle.}
    \label{fig:measurement-dates}
\end{figure}

The independent variable of this RQ is the temporal proximity of TLP. This variable has three treatments, i.e., proximity points, which reflect the lifecycle of project development (as shown in \autoref{fig:measurement-dates}):
\begin{itemize}
    \item Before Ticket Assignment, aka, \textbf{Open}: The ticket is created and not yet assigned. We measure this proximity point as one second before the ticket assignment date, as reported in JIRA.
    \item Before First Commit, aka, \textbf{InProgress}: The ticket is assigned and no commit has been submitted yet. We measure this proximity point as one second before the first commit, as reported in Git.  
    \item After Last Commit, aka, \textbf{Closed}: The ticket is completely implemented. We measure this proximity point as one second after the last commit, as reported in Git. 
\end{itemize}

% issue lifecycle diagram

\subsubsection{Dependent Variables}
\label{sec:design_rq1_dv}

The main dependent variable of this RQ is the accuracy of TLP. As accuracy indicators of TLP, we used the following six metrics, which are standards in machine learning \citep{DBLP:journals/ese/PatelAH24}:
\begin{itemize}
    \item AUC: aka, Area Under the Receiving Operating Characteristic Curve \citep{DBLP:journals/ese/FalessiLCEC23}, is the area under the curve of true positive rate versus false positive rate, which is defined by setting multiple thresholds. A positive instance is a bug-inducing ticket, whereas a negative instance is a non-bug-inducing ticket. AUC has the advantage of being threshold independent and, therefore, it is recommended for evaluating defect prediction techniques when, like in our TLP context, the costs associated with misclassification errors are unknown or not easily comparable \citep{DBLP:journals/tse/LessmannBMP08};
    \item Precision: It is an accuracy metric describing how often the positive predictions of the model are correct \citep{DBLP:journals/tse/FalessiRGC20}; 
    \item Recall: It describes how often the model predicts true positive entities as actually positive \citep{DBLP:journals/tse/FalessiRGC20}. 
    \item Kappa: It describes how well the predictor performs compared against a random guesser \citep{DBLP:journals/tosem/FalessiAP22};
    \item Specificity: It is a concept close to Recall but related to the negative class. It describes how often the model predicts negative entities as actually negative \citep{DBLP:journals/ese/PatelAH24};
    \item GMean: When used to evaluate models performing binary classification tasks, it is defined as the square root of the product between recall and specificity \citep{DBLP:journals/ese/PatelAH24}.
\end{itemize}

\subsubsection{Experimental Setup}
%\label{sec:design_rq2_analysis}
To reduce variability in the experimental methodology, we took inspiration from what \citet{DBLP:journals/ese/PatelAH24} did in JIT context. 

Specifically, as the validation technique \citep{DBLP:journals/corr/abs-1809-01510}, we used the same sliding window by \citet{DBLP:journals/ese/PatelAH24}. We initialize the window for each dataset, taking a batch of the first 1000 instances and split it into 80\% training and 20\% testing. After each iteration, we update the window by removing the first 200 instances and adding the next 200 in the dataset. The procedure described up to now is repeated until all instances in the dataset have been consumed.

Regarding supervised machine learning models, we used the same three used by \citet{DBLP:journals/ese/PatelAH24}:
\begin{itemize}
    \item Random Forest (RF): This model is an ensemble learning method that consults a random subset of decision trees whenever it predicts to reduce correlation among the bagged trees \citep{james2023introduction};
    \item Logistic Regression (LR): It is a variant of the linear regression model specialized in binary classification tasks \citep{james2023introduction};
    \item Neural Network (NN): It is a model inspired by the human brain that can learn complex patterns in the data by fitting the weights of the connections between its neurons, which are organized in layers and exchange information through activation functions \citep{james2023introduction};
\end{itemize}

Regarding the feature selection technique, we evaluated the classifiers with and without feature selection. In particular, we considered the filter approach, which is a model-independent approach that builds a subset of features by selecting the highly correlated ones with the target and has little correlation among them \citep{witten2002data}. However, feature selection could be counterproductive \citep{ZHAO2021106652, osman2018impact, HUANG2024111968} due to many reasons such as the possible elimination of useful information or under-fitting.

Regarding balancing, we performed the evaluation without balancing and with the SMOTE technique \citep{DBLP:journals/corr/abs-1106-1813}. Balancing the population numbers of bug-inducing and non-bug-inducing tickets can help the models learn better from the dataset. However, since our datasets are unbalanced towards the positives, i.e., bug-inducing tickets are the majority, the effects of balancing is likely counterproductive.

%\begin{figure}
%    \centering
%    \includegraphics[width=\linewidth]{imgs/design/diagrams/80-20.drawio.png}
%    \caption{80-20 Ordered Holdout example using the first commit date as measurement date.}
%    \label{fig:80-20}
%\end{figure}

%\begin{figure}
%    \centering
%    \includegraphics[width=\linewidth]{imgs/design/diagrams/sliding-window.drawio.png}
%    \caption{Sliding Window example using the first commit date as measurement date.}
%    \label{fig:sliding-window}
%\end{figure}

\subsubsection{Hypothesis and testing}
\label{sec:design_rq1_hypotheses}
Our null hypothesis H01 is that the accuracy of TLP does not vary across temporal proximity points.

To assess the statistical significance of differences between the treatments, we employed the Friedman test, a non-parametric test suitable for paired data \citep{Friedman1937}. This test ranks the data for each subject across conditions, thereby mitigating the influence of non-normal distributions. The Friedman test efficiently handles the within-subject correlations since each sliding window provides measurements for each of the three proximity points. We tested each accuracy metric for significant differences across treatments within the same dataset and classifier. We set alpha to 0.05 \citep{gibbons2014nonparametric}.

In the context of the Friedman test, a commonly recommended measure of effect size is Kendall’s W \citep{Kendall1939}. It represents the degree of agreement in the rankings across the conditions and is particularly well-suited for repeated measures designs with more than two conditions. Kendall’s W ranges from 0 (no agreement) to 1 (complete agreement). In our experimental context, where each window provides measurements under three different proximity points, Kendall’s W offers a robust indicator, complementing the Friedman test's chi-square statistic.

%%%% RQ2
\subsection{RQ2: \rqtwo}
\label{sec:design_rq2}

\subsubsection{Introduction}
\label{sec:design_rq2_intro}
Some TLP features can be more or less available and more or less accurate than others at different proximity points. For instance, JIT features are available only when the ticket is Closed and not available when it is Open. Features related to the assigned developer are available when in progress, and can change in Closed. Moreover, the more features we use, the more data we need to train our models. This problem is known in the Machine Learning landscape as the Curse of Dimensionality \citep{crespo2022curse}. For the above reasons, a project manager using TLP could be interested in which features are easier to measure and most informative at a specific proximity point. This RQ explores how the importance of each feature varies with proximity.  

\subsubsection{Independent Variables}
\label{sec:design_rq2_iv}

The independent variables of this RQ are: 
\begin{enumerate}
\item the temporal proximity of TLP. This has three treatments as in RQ1: Open, In Progress, and Closed.
\item the features used for TLP, which are the 72 detailed in \autoref{sec:features} and grouped in seven families: Code, Developer, External Temperature, Internal Temperature, Intrinsic, Ticket to Tickets, and JIT.
\end{enumerate}

\subsubsection{Dependent Variables}
\label{sec:design_rq2_dv}
The dependent variable is the prediction power of features. We used the  Information Gain Ratio (IGR) to measure the prediction power, as we successfully used in a similar work  \citet{DBLP:journals/tse/FalessiRGC20}. One of the advantages of IGR over correlation-based metrics like Spearman is that it is agnostic to classifiers and prediction accuracy metrics. However, the downside is that it is hard to interpret without a given reference \citep{DBLP:journals/tse/FalessiRGC20}. Finally, it is important to note that IGR evaluates a feature’s predictive power independently, without considering interactions with other features. However, when a prediction model uses multiple features, as in TLP, a feature with low IGR can be more beneficial than a feature with high IGR if it exhibits lower redundancy with the existing feature set. 
We computed the IGR for all features and examined their distribution across feature families; this allows us to understand how entire families perform in different proximity points. Afterwards, we ranked each feature based on IGR for each project, identifying the top and bottom 10 features in each proximity point. This approach allows us to observe which features are most or least informative and how their importance varies across projects and proximity point.

\subsubsection{Hypotheses and testing}
\label{sec:design_rq2_hypotheses}

Our null hypothesis, H20, is that the power of TLP features does not vary across feature family, temporal locality, and their interaction.
As a statistical test, we use the same as RQ1.

\subsubsection{Experimental Setup}
To produce the evaluation results, we applied the same RQ1 setup. Thus, for each window, we computed the IGR of each feature.

\subsection{Measurement Procedure}
\label{sec:design_datasets}

%\begin{figure}
    %\centering
    %\includegraphics[width=\linewidth]{imgs/design/diagrams/phases-overview.drawio.png}
    %\caption{Measurement procedure.}
   % \label{fig:phases_overview}
%\end{figure}

To create TLP datasets instead of reinventing the wheel and creating a TLP dataset from scratch, we searched for available datasets that could be used or adapted as TLP datasets. Specifically, the TLP dataset requires a ticket to be labelled as bug-inducing or not-bug-inducing. Since labelling entities as buggy or not is a line of research by itself, we aimed at reusing trustable labels. However, we found no dataset labelling tickets as bug-inducing or not-bug-inducing. According to our definition of bug-inducing-ticket, a ticket is bug-inducing if, and only if, at least a commit implementing the ticket is buggy. Therefore, we continued our search by looking at JIT datasets providing the buggy/not-buggy labels for commits and we found three datasets related to three publications \citep{DBLP:journals/ese/CabralMOPT23, DBLP:journals/ese/FalessiLCEC23, DBLP:conf/msr/KeshavarzN22}.

To select the projects in these datasets, we ranked the projects according to the linkage proportion of buggy tickets, as shown in \autoref{table:Join}. Note that we filtered anomalous tickets since they can confuse the models. In order to remove anomalous tickets, we applied the following filters:
\begin{itemize}
    \item \texttt{ExclusiveBuggyCommitsOnly}: we remove a ticket if all its buggy commits are related to other tickets. This is an important filter since, in this case, we cannot establish which ticket, of the ones related to the buggy commit, induced the bug.
    \item \texttt{FirstCommitAfterOpeningDate}: we remove a ticket if the first commit date is after its opening date. This is important because we use the date of the first commit to measure the temporal proximity.
    \item \texttt{ClearRepository}: we remove a ticket if it belongs to a project for which it has been impossible to establish the main repository. 
    \item \texttt{NoSnoring}:  we remove a ticket if within the last 20\% of the tickets. This is important  to mitigate the impact of snoring \citep{DBLP:journals/tosem/FalessiAP22}. Note that this was not required for the LeveragingJIT dataset since it was already applied while building the original  dataset \citep{DBLP:journals/ese/FalessiLCEC23};
    \item \texttt{CommitAfterOpeningDate}: We removed a ticket if the date of its first commit is before its Opening Date.
\end{itemize}

% \usepackage{lscape}
%\begin{landscape}
    \begin{table}[]
    \centering
    \caption{List of projects considered in the study, ranked by the percentage of buggy
    commits linked to Jira tickets. Each project is identified by its name and the
    datasets its commits come from. The usable ticket count is the number of tickets that
    have survived the filtering process.}
    \label{table:Join}
    \begin{tabular}{lccccc}
    \hline
    \multicolumn{1}{c}{\textbf{Dataset}} & \textbf{Project}   & \textbf{\% Buggy Linkage} & \textbf{\% Linkage} & \textbf{\# Tickets} & \textbf{\# Usable Tickets} \\ \hline
    leveragingjit                        & ZooKeeper          & 100                       & 60                  & 91                  & 91                         \\
    apachejit                            & Hive               & 99                        & 99                  & 6443                & 5025                       \\
    apachejit                            & HBase              & 97                        & 94                  & 7128                & 5402                       \\
    leveragingjit                        & Tika               & 96                        & 93                  & 126                 & 124                        \\
    apachejit                            & Spark              & 94                        & 86                  & 1298                & 631                        \\
    apachejit                            & ZooKeeper          & 93                        & 92                  & 748                 & 559                        \\
    apachejit                            & Kafka              & 83                        & 61                  & 1340                & 600                        \\
    apachejit                            & Camel              & 83                        & 62                  & 7716                & 6039                       \\
    apachejit                            & Cassandra          & 81                        & 63                  & 4078                & 3163                       \\
    apachejit                            & Flink              & 78                        & 48                  & 4265                & 3295                       \\
    apachejit                            & ActiveMQ Classic   & 74                        & 53                  & 2195                & 1654                       \\
    apachejit                            & Ignite             & 73                        & 49                  & 2870                & 2276                       \\
    apachejit                            & Zeppelin           & 73                        & 61                  & 866                 & 291                        \\
    jitsdp                               & Camel              & 67                        & 55                  & 9095                & 7103                       \\
    leveragingjit                        & ActiveMQ Artemis   & 61                        & 30                  & 133                 & 103                        \\
    leveragingjit                        & Qpid               & 59                        & 47                  & 478                 & 394                        \\
    apachejit                            & Groovy             & 57                        & 39                  & 2612                & 2004                       \\
    leveragingjit                        & Directory ApacheDS & 50                        & 15                  & 44                  & 44                         \\
    leveragingjit                        & Maven              & 32                        & 18                  & 255                 & 240                        \\
    leveragingjit                        & Nutch              & 29                        & 24                  & 44                  & 44                         \\
    leveragingjit                        & OpenJPA            & 21                        & 17                  & 69                  & 66                         \\
    leveragingjit                        & Groovy             & 20                        & 17                  & 116                 & 114                        \\
    apachejit                            & Hadoop Map/Reduce  & 17                        & 39                  & 833                 & 661                        \\
    apachejit                            & Hadoop HDFS        & 7                         & 20                  & 2842                & 2236                       \\
    apachejit                            & Hadoop Common      & 0                         & 99                  & 3249                & 2586                       \\ \hline
    \end{tabular}
    \end{table}
   % \end{landscape}

We concluded the search by selecting the projects HIVE and HBASE from apachejit since they have the highest buggy linkage and have lots of usable tickets. Afterwards, we created our TLP dataset by measuring the features described in \autoref{sec:features}, producing a dataset for each project and temporal proximity point combination, resulting in six datasets. The tickets in each dataset are ordered by the date of the specific proximity point; this is to avoid during prediction, future information is used to predict the past \citep{DBLP:journals/corr/abs-1809-01510}.

\section{Results}\label{sec:results}
%% risultati devono essere concisi e compartimentati
%% Figure reports/displays/shows …. 

%%%% RQ1
In the absence of established guidelines for setting up experiments in TLP, it is first necessary to determine the experimental configuration that yields the highest accuracy before addressing the research questions. Our preliminary analysis, presented in Section A1 of the Appendix, indicates that feature selection and data balancing do not improve the AUC performance in TLP. Consequently, we answer our research questions based on experiments conducted without applying feature selection or balancing techniques.

\subsection{RQ1: \rqone}
\label{sec:res_rq1}

%%%%%%%%%% SW %%%%%%%%%%

\begin{figure}
    \centering
    \includegraphics[width=\textwidth]{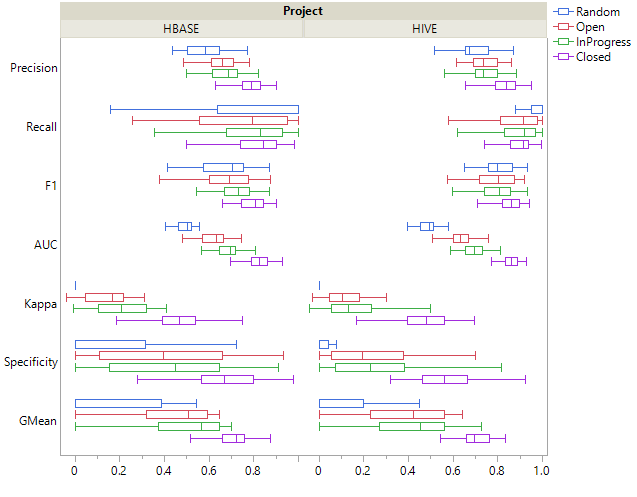}
    % NEW ->RQ1/SW/AccuracyResultsReduced - Graph Builder 2
    \caption{Distributions of TLP accuracy in three proximity points.}
    \label{fig:rq1_SW_accuracy}
\end{figure}

\autoref{fig:rq1_SW_accuracy} shows the distributions of TLP accuracy in three proximity points and a random approach. \autoref{tab:rq1_gain} synthesizes \autoref{fig:rq1_SW_accuracy} by reporting, for each proximity point, the average improvement in TLP accuracy, averaged across our three classifiers, relative to the random baseline.

% OLD \input{Tables/results/rq1/SW/gain_hbase.tex}
% OLD \input{Tables/results/rq1/SW/gain_hive.tex}
%->RQ1/SW/gain.xlsx
% \usepackage{multirow}
\begin{table}[]
    \centering
    \caption{Average gain across classifiers in TLP accuracy using SW in HBASE and HIVE compared to random TLP.}
    \label{tab:rq1_gain}

    \begin{tabular}{ccccccccc}
    \toprule
    %\multicolumn{9}{c}{\textbf{Gain on random}}  \\ \midrule
    \textbf{Projects}  & \textbf{Proximity Point} & \textbf{Precision} & \textbf{Recall} & \textbf{F1} & \textbf{AUC} & \textbf{Kappa*} & \textbf{Specificity} & \textbf{GMean}  \\ \midrule
                       & Open                     & 20\%               &  -7\%           & 7\%         & 25\%         & 14\%            & 98\%                 & 163\%           \\ 
    \textbf{HBASE}     & InProgress               & 23\%               &  -4\%           & 10\%        & 37\%         & 20\%            & 117\%                & 193\%           \\ 
                       & Closed                   & 44\%               &   0\%           & 22\%        & 67\%         & 46\%            & 228\%                & 314\%           \\ \midrule
                       & Open                     & 5\%                & -10\%           & -3\%        & 30\%         & 12\%            & 538\%                & 311\%           \\ 
    \textbf{HIVE}      & InProgress               & 6\%                &  -8\%           & -1\%        & 43\%         & 15\%            & 589\%                & 364\%           \\ 
                       & Closed                   & 19\%               &  -7\%           & 6\%         & 76\%         & 47\%            & 1377\%               & 663\%           \\ \bottomrule
    \end{tabular}
    \end{table}

\autoref{tab:rq1_friedman_test} reports the statistical test results comparing the accuracy of the same classifier on the same project over different proximity points. According to \autoref{tab:rq1_friedman_test} we can reject H01 in all accuracy metrics other than Recall and hence we can claim that the proximity impacts TLP accuracy.

%\input{Tables/results/rq1/SW/statistical_test.tex}
% NEW ->RQ1/SW/FriedmanTestResults.xlsx
% \usepackage{multirow}
\begin{table}[]
    \centering
    \caption{Statistical test comparison on the impact on IGR of Feature Family, Proximity and their interaction using sliding-window.}
    \label{tab:rq1_friedman_test}
    \begin{tabular}{clccccccc}
    \toprule
    
    \textbf{Projects}  & \textbf{Statistical Test}  & \textbf{Precision} & \textbf{Recall} & \textbf{F1} & \textbf{AUC} & \textbf{Kappa} & \textbf{Specificity} & \textbf{GMean} \\ \midrule
    \textbf{HIVE}      & Pvalue            & 0,0000             & 0,0863          & 0,0000      & 0,0000       & 0,0000         & 0,0000               & 0,0000         \\ 
                       & KendallsW         & 0,6775             & 0,1225          & 0,6925      & 0,8125       & 0,7500         & 0,6100               & 0,7525         \\ \midrule
    \textbf{HBASE}     & Pvalue            & 0,0000             & 0,6146          & 0,0028      & 0,0000       & 0,0000         & 0,0001               & 0,0000         \\ 
                       & KendallsW         & 0,6842             & 0,0256          & 0,3102      & 0,9058       & 0,7749         & 0,5014               & 0,6011         \\ \bottomrule
    \end{tabular}
    \end{table}

\subsection{RQ2: \rqtwo}
\label{sec:res_rq2}

%Given that in RQ1 we observed a higher accuracy in SW than 80-20, RQ2 results are based on SW only.

\begin{figure}
    \centering
    \includegraphics[width=\textwidth]{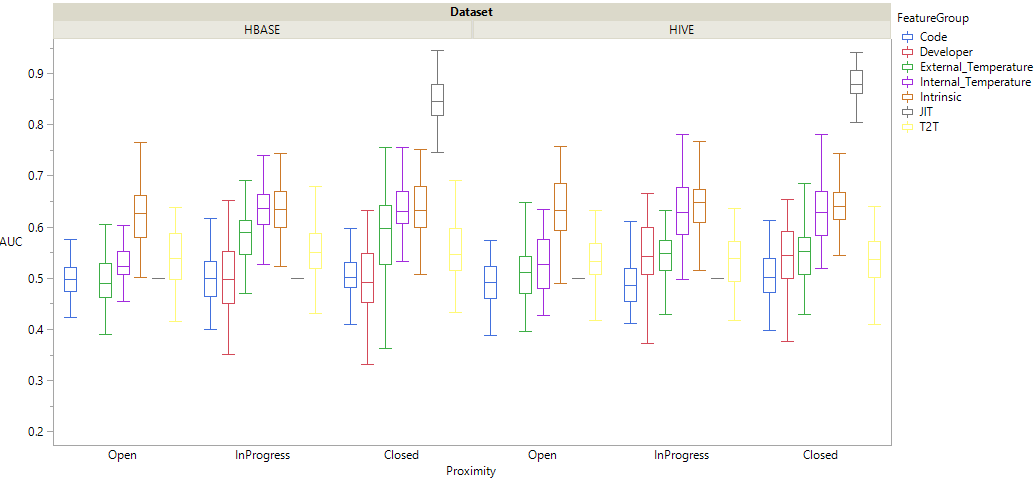}
    % NEW ->Rq2/MW/AccuracyByGroupMW.png
    \caption{ Distributions of TLP accuracy in terms of AUC achieved by a single feature family,  in different proximity points and projects, across windows. }
    \label{fig:rq2_SW_AccuracybyGroup}
\end{figure}

\autoref{fig:rq2_SW_AccuracybyGroup} shows, for each feature family, the distribution of TLP accuracy, in terms of AUC, across sliding windows, by proximity point and project.

\begin{figure}
    \centering
    \includegraphics[width=\textwidth]{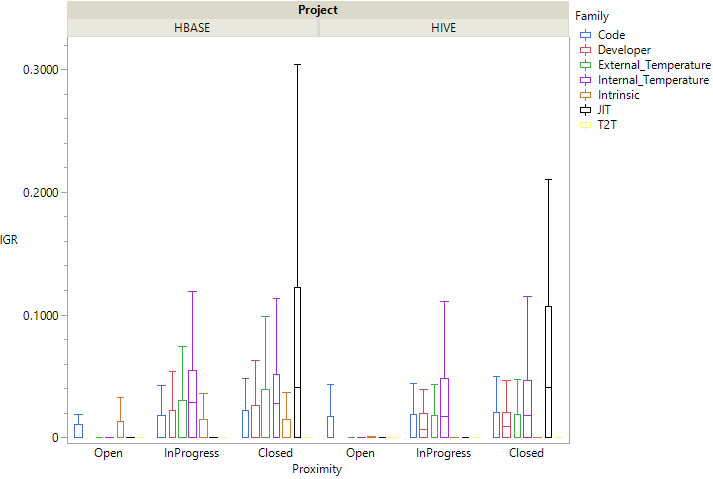}
    % NEW -> MW/FeatureSelectionResults - Graph Builder.png
    \caption{ Distributions of feature family power,  in different proximity points and projects, in terms of max and mean IGR,  and max and mean selection, across windows. }
    \label{fig:rq2_SW_igr}
\end{figure}

\autoref{fig:rq2_SW_igr} shows, for each feature family, the distribution of IGR, across sliding windows, by proximity point and project.

%\autoref{tab:rq2-mw-fs-results} reports the summary of feature selection results.  

\autoref{tab:rq2-mw-stat-tests} reports the Statistical test comparison on the impact on IGR of Feature Family, Proximity and their interaction. According to \autoref{tab:rq2-mw-stat-tests}, we can reject H20 in both projects, and hence we can claim that the prediction power of features varies according to feature family, proximity point, and their interaction.

Finally, \autoref{tab:top_10_igr} reports the top 10 features out of 72, ranked by IGR in a specific project in a specific proximity point. Similarly, \autoref{tab:bottom_10_igr} reports bottom top 10 features out of 72, ranked by IGR in a specific project in a specific proximity point.

% \scriptsize
% \input{Tables/results/rq2/SW/fs-results.tex}
% NEW ->Rq2/MW/Summary of FeatureSelectionResults grouped by Project, Proximity, Feature, Family.xlsx
% \input{Tables/results/rq2/feature_selection.tex}
\normalsize

% \input{Tables/results/rq2/SW/rq2-SW-stats-tests.tex}
% \usepackage{booktabs}
% \usepackage{multirow}
\begin{table}[]
    \centering
    \caption{Statistical test comparison on the impact on IGR of Feature Family, Proximity and their interaction using moving-window. }
    \label{tab:rq2-mw-stat-tests}
    \begin{tabular}{@{}lrr@{}}
    \toprule
    \multicolumn{1}{c}{\multirow{2}{*}{\textbf{Independent Variable}}} & \multicolumn{2}{c}{\textbf{Pvalue}}                                    \\
    \multicolumn{1}{c}{}                                               & \multicolumn{1}{c}{\textbf{HBASE}} & \multicolumn{1}{c}{\textbf{HIVE}} \\ \midrule
    FeatureFamily                                                      & 0.0001                             & 0.0001                            \\
    Proximity                                                          & 0.0001                             & 0.0001                            \\
    $\text{Proximity} \times \text{FeatureFamily} $                                           & 0.0001                             & 0.0001                            \\ \bottomrule
    \end{tabular}
    \end{table}
%->ok

\begin{table}[h!]
    \centering
    \caption{Top 10 features by IGR per project, grouped by proximity point.}
    \label{tab:top_10_igr}
    % OPEN
    \begin{subfigure}[b]{0.96\textwidth}
        \begin{minipage}[t]{\linewidth} % Use linewidth within minipage
            \centering
            \scriptsize
            \begin{tabularx}{\linewidth}{lXrrrlXrr}
    % \toprule
    \multicolumn{4}{c}{\textbf{HBASE - Open}} & & \multicolumn{4}{c}{\textbf{HIVE - Open}} \\
    \cmidrule(lr){1-4} \cmidrule(lr){6-9}
    %\textbf{FF} & \textbf{FN} & \textbf{Mean(IGR)} & \textbf{OR} & & \textbf{FF} & \textbf{FN} & \textbf{Mean(IGR)} & \textbf{OR} \\
    \textbf{Feature Family} & \textbf{Feature Name} & \textbf{Mean(IGR)} & \textbf{Rank}  & & \textbf{Feature Family} & \textbf{Feature Name} & \textbf{Mean(IGR)} & \textbf{Rank} \\
    \cmidrule(lr){1-4} \cmidrule(lr){6-9}
    % \midrule[\heavyrulewidth] % Optional: Use a thicker midrule for visual separation
    I     & \textit{type}                             & 0,042090 & 19 & & I      & \textit{type}                             & 0,049482 & 14\\ % & 1  \\
    C     & \textit{code\_size-number\_of\_languages} & 0,021765 & 33 & & I      & \textit{priority}                         & 0,017441 & 30\\ % & 2  \\
    I     & \textit{priority}                         & 0,017758 & 41 & & C      & \textit{code\_size-number\_of\_languages} & 0,014348 & 38\\ % & 3  \\
    C     & \textit{code\_quality-smells\_count}      & 0,017296 & 44 & & I      & \textit{components-max\_bugginess}        & 0,013375 & 41\\ % & 4  \\
    C     & \textit{code\_size-total\_LOCs}           & 0,016226 & 45 & & C      & \textit{code\_size-total\_LOCs}           & 0,009943 & 45\\ % & 5  \\
    E\_T  & \textit{temporal\_locality}               & 0,015672 & 46 & & C      & \textit{code\_size-number\_of\_files}     & 0,008898 & 48\\ % & 6  \\
    E\_T  & \textit{temporal\_locality-weighted}      & 0,015567 & 47 & & I      & \textit{nlp4re\_description-DA\_ACT}      & 0,007655 & 54\\ % & 7  \\
    I     & \textit{nlp4re\_description-EX\_RDS}      & 0,014735 & 51 & & I\_T   & \textit{issue\_part-count}                & 0,007394 & 57\\ % & 8  \\
    C     & \textit{code\_size-number\_of\_files}     & 0,011611 & 60 & & I      & \textit{nlp4re\_description-EX\_VRB}      & 0,006787 & 59\\ % & 9  \\
    I     & \textit{nlp4re\_description-DA\_ACT}      & 0,010101 & 65 & & E\_T   & \textit{temporal\_locality-weighted}      & 0,006778 & 60\\ % & 10 \\
    \cmidrule(lr){1-4} \cmidrule(lr){6-9}
\end{tabularx}
            % \caption{Top 10 - Open}
        \end{minipage}
        %\label{tab:table_bottom_hive_open}
    \end{subfigure}
    \vskip\baselineskip
    % IN PROGRESS
    \begin{subfigure}[b]{0.96\textwidth}
        \begin{minipage}[t]{\linewidth} % Use linewidth within minipage
            \centering
            \scriptsize
            \begin{tabularx}{\linewidth}{lXrrrlXrr}
    % \toprule
    \multicolumn{4}{c}{\textbf{HBASE - In Progress}} & & \multicolumn{4}{c}{\textbf{HIVE - In Progress}} \\
    \cmidrule(lr){1-4} \cmidrule(lr){6-9}
    %\textbf{FF} & \textbf{FN} & \textbf{Mean(IGR)} & \textbf{OR} & & \textbf{FF} & \textbf{FN} & \textbf{Mean(IGR)} & \textbf{OR} \\
    \textbf{Feature Family} & \textbf{Feature Name} & \textbf{Mean(IGR)} & \textbf{Rank}  & & \textbf{Feature Family} & \textbf{Feature Name} & \textbf{Mean(IGR)} & \textbf{Rank} \\
    \cmidrule(lr){1-4} \cmidrule(lr){6-9}
    % \midrule[\heavyrulewidth] % Optional: Use a thicker midrule for visual separation
    I\_T  & \textit{activities-count}                   & 0,063784  &  7 & & I\_T  & \textit{activities-histories}               & 0,060274 & 7  \\ % & 1  \\
    I\_T  & \textit{activities-comments}                & 0,062051  &  8 & & I\_T  & \textit{nlp4re\_sentiment-CM\_NNS}          & 0,058154 & 8  \\ % & 2  \\
    I\_T  & \textit{activities-histories}               & 0,049484  & 15 & & I\_T  & \textit{activities-count}                   & 0,053487 & 11 \\ % & 3  \\
    I\_T  & \textit{nlp4re\_sentiment-CM\_NNS}          & 0,042883  & 17 & & I     & \textit{type}                               & 0,051830 & 12 \\ % & 4  \\
    I\_T  & \textit{issue\_participants-count}          & 0,042722  & 18 & & I\_T  & \textit{activities-comments}                & 0,043017 & 15 \\ % & 5  \\
    I     & \textit{type}                               & 0,040167  & 20 & & I\_T  & \textit{issue\_part-count}                  & 0,032542 & 20 \\ % & 6  \\
    E\_T  & \textit{commits\_while\_in\_progress-count} & 0,036187  & 25 & & E\_T  & \textit{commits\_while\_in\_progress-count} & 0,024242 & 23 \\ % & 7  \\
    E\_T  & \textit{commits\_while\_in\_progress-churn} & 0,035565  & 26 & & I     & \textit{priority}                           & 0,019878 & 28 \\ % & 8  \\
    I\_T  & \textit{nlp4re\_sentiment-CM\_PNS}          & 0,031066  & 28 & & E\_T  & \textit{commits\_while\_in\_progress-churn} & 0,016556 & 32 \\ % & 9  \\
    I     & \textit{priority}                           & 0,024077  & 29 & & I     & \textit{components-max\_bugginess}          & 0,015509 & 34 \\ % & 10 \\
    \cmidrule(lr){1-4} \cmidrule(lr){6-9}
\end{tabularx}
            % \caption{Top 10 - In Progress}
        \end{minipage}
        %\label{tab:table_bottom_hive_open}
    \end{subfigure}
    \vskip\baselineskip
    % CLOSED
    \begin{subfigure}[b]{0.96\textwidth}
        \begin{minipage}[t]{\linewidth} % Use linewidth within minipage
            \centering
            \scriptsize
            \begin{tabularx}{\linewidth}{lXrrrlXrr}
    % \toprule
    \multicolumn{4}{c}{\textbf{HBASE - Closed}} & & \multicolumn{4}{c}{\textbf{HIVE - Closed}} \\
    \cmidrule(lr){1-4} \cmidrule(lr){6-9}
    %\multicolumn{2}{l}{\hspace{0.5cm}\textbf{Feature} } \\ 
    % \textbf{FF} & \textbf{FN} & \textbf{Mean(IGR)} & \textbf{OR} & & \textbf{FF} & \textbf{FN} & \textbf{Mean(IGR)} & \textbf{OR} \\
    \textbf{Feature Family} & \textbf{Feature Name} & \textbf{Mean(IGR)} & \textbf{Rank}  & & \textbf{Feature Family} & \textbf{Feature Name} & \textbf{Mean(IGR)} & \textbf{Rank} \\
    \cmidrule(lr){1-4} \cmidrule(lr){6-9}
    % \midrule[\heavyrulewidth] % Optional: Use a thicker midrule for visual separation
    JIT     & \textit{jit-la-SUM}                          & 0,233243 & 1  & & JIT   & \textit{jit-la-SUM}                & 0,157454  & 1  \\ % & 1  \\
    JIT     & \textit{jit-nf-MAX}                          & 0,170639 & 2  & & JIT   & \textit{jit-nf-MAX}                & 0,142956  & 2  \\ % & 2  \\
    JIT     & \textit{jit-ent-MAX}                         & 0,148063 & 3  & & JIT   & \textit{jit-nd-MAX}                & 0,137040  & 3  \\ % & 3  \\
    JIT     & \textit{jit-ld-SUM}                          & 0,128697 & 4  & & JIT   & \textit{jit-ent-MAX}               & 0,113708  & 4  \\ % & 4  \\
    JIT     & \textit{jit-nd-MAX}                          & 0,126211 & 5  & & JIT   & \textit{jit-ld-SUM}                & 0,111820  & 5  \\ % & 5  \\
    JIT     & \textit{jit-ns-MAX}                          & 0,068924 & 6  & & JIT   & \textit{jit-ns-MAX}                & 0,071429  & 6  \\ % & 6  \\
    I\_T    & \textit{activities-count}                    & 0,060702 & 9  & & I\_T  & \textit{activities-histories}      & 0,056970  & 9  \\ % & 7  \\
    I\_T    & \textit{activities-comments}                 & 0,058418 & 10 & & I\_T  & \textit{nlp4re\_sentiment-CM\_NNS} & 0,055771  & 10  \\ % & 8  \\
    E\_T    & \textit{commits\_while\_in\_progress-churn}  & 0,057017 & 11 & & I\_T  & \textit{activities-count}          & 0,051782  & 13  \\ % & 9  \\
    JIT     & \textit{jit-ndev-MAX}                        & 0,054313 & 12 & & I     & \textit{type}                      & 0,042701  & 16  \\ % & 10 \\
    \cmidrule(lr){1-4} \cmidrule(lr){6-9}
\end{tabularx}
            % \caption{Top 10 - Closed}
        \end{minipage}
        %\label{tab:table_bottom_hive_open}
    \end{subfigure}
\end{table}
\begin{table}[h!]
    \centering
    \caption{Bottom 10 features by IGR per project, grouped by proximity point.}
    \label{tab:bottom_10_igr}
    % OPEN
    \begin{subfigure}[b]{0.96\textwidth}
        \begin{minipage}[t]{\linewidth} % Use linewidth within minipage
            \centering
            \scriptsize
            \begin{tabularx}{\linewidth}{lXrrrlXrr}
    % \toprule
    \multicolumn{4}{c}{\textbf{HBASE - Open}} & & \multicolumn{4}{c}{\textbf{HIVE - Open}} \\
    \cmidrule(lr){1-4} \cmidrule(lr){6-9}
    % \textbf{FF} & \textbf{FN} & \textbf{Mean(IGR)} & \textbf{OR} & & \textbf{FF} & \textbf{FN} & \textbf{Mean(IGR)} & \textbf{OR} \\
    \textbf{Feature Family} & \textbf{Feature Name} & \textbf{Mean(IGR)} & \textbf{Rank}  & & \textbf{Feature Family} & \textbf{Feature Name} & \textbf{Mean(IGR)} & \textbf{Rank} \\
    \cmidrule(lr){1-4} \cmidrule(lr){6-9}
    % \midrule[\heavyrulewidth] % Optional: Use a thicker midrule for visual separation
    I    & \textit{nlp4re\_description-DA\_WKP}                                     & 0,000579 & 159 & & I     & \textit{nlp4re\_description-EX\_ICP}                       & 0,000354 & 149 \\
    T2T  & \textit{buggy\_similarity-avg\_similarity\_tfidf\_cosine\_text}          & 0,000438 & 161 & & I\_T  & \textit{activities-comments}                               & 0,000260 & 151 \\
    T2T  & \textit{buggy\_similarity-avg\_similarity\_euclidean\_distance\_title}   & 0,000366 & 163 & & I     & \textit{nlp4re\_description-DA\_CNT}                       & 0,000142 & 153 \\
    I    & \textit{components-max\_bugginess}                                       & 0,000343 & 164 & & I     & \textit{nlp4re\_description-DA\_INC}                       & 0,000043 & 157 \\
    I\_T & \textit{activities-work\_items\_count}                                   & 0,000000 & 169 & & I\_T  & \textit{activities-work\_items\_count}                     & 0,000000 & 159 \\
    I\_T & \textit{nlp4re\_sentiment-IT\_SUB}                                       & 0,000000 & 170 & & I\_T  & \textit{nlp4re\_sentiment-CM\_PNS}                         & 0,000000 & 160 \\
    I    & \textit{nlp4re\_description-DA\_CNT}                                     & 0,000000 & 171 & & I     & \textit{nlp4re\_description-DA\_SRC}                       & 0,000000 & 161 \\
    I    & \textit{nlp4re\_description-DA\_INC}                                     & 0,000000 & 172 & & I     & \textit{nlp4re\_description-EX\_ENT}                       & 0,000000 & 162 \\
    I    & \textit{nlp4re\_description-DA\_SRC}                                     & 0,000000 & 173 & & I     & \textit{nlp4re\_description-EX\_RDS}                       & 0,000000 & 163 \\
    T2T  & \textit{buggy\_similarity-max\_similarity\_tfidf\_cosine\_text}          & 0,000000 & 175 & & T2T   & \textit{buggy\_similarity-max\_similarity\_jaccard\_text}  & 0,000000 & 165 \\
    \cmidrule(lr){1-4} \cmidrule(lr){6-9}
\end{tabularx}

            % \caption{Bottom 10 - Open}
        \end{minipage}
        %\label{tab:table_bottom_hive_open}
    \end{subfigure}

    \vskip\baselineskip
    % IN PROGRESS
    \begin{subfigure}[b]{0.96\textwidth}
        \begin{minipage}[t]{\linewidth} % Use linewidth within minipage
            \centering
            \scriptsize
            \begin{tabularx}{\linewidth}{lXrrrlXrr}
    % \toprule
    \multicolumn{4}{c}{\textbf{HBASE - In Progress}} & & \multicolumn{4}{c}{\textbf{HIVE - In Progress}} \\
    \cmidrule(lr){1-4} \cmidrule(lr){6-9}
    % \textbf{FF} & \textbf{FN} & \textbf{Mean(IGR)} & \textbf{OR} & & \textbf{FF} & \textbf{FN} & \textbf{Mean(IGR)} & \textbf{OR} \\
    \textbf{Feature Family} & \textbf{Feature Name} & \textbf{Mean(IGR)} & \textbf{Rank}  & & \textbf{Feature Family} & \textbf{Feature Name} & \textbf{Mean(IGR)} & \textbf{Rank} \\
    \cmidrule(lr){1-4} \cmidrule(lr){6-9}
    % \midrule[\heavyrulewidth] % Optional: Use a thicker midrule for visual separation
    T2T     & \textit{buggy\_similarity-max\_similarity\_tfidf\_cosine\_title}       & 0,001537 & 147 & & I     & \textit{nlp4re\_description-DA\_INC}                      & 0,000043 & 155 \\
    T2T     & \textit{buggy\_similarity-max\_similarity\_tfidf\_cosine\_text}        & 0,000915 & 152 & & E\_T  & \textit{latest\_commit-churn}                             & 0,000000 & 166 \\
    E\_T    & \textit{latest\_commit-churn}                                          & 0,000596 & 158 & & I\_T  & \textit{activities-work\_items\_count}                    & 0,000000 & 167 \\
    I\_T    & \textit{nlp4re\_sentiment-IT\_SUB}                                     & 0,000556 & 160 & & I     & \textit{nlp4re\_description-DA\_CNT}                      & 0,000000 & 168 \\
    I       & \textit{nlp4re\_description-DA\_WKP}                                   & 0,000191 & 166 & & I     & \textit{nlp4re\_description-DA\_SRC}                      & 0,000000 & 169 \\
    I\_T    & \textit{activities-work\_items\_count}                                 & 0,000000 & 176 & & I     & \textit{nlp4re\_description-DA\_WKP}                      & 0,000000 & 170 \\
    I       & \textit{nlp4re\_description-DA\_CNT}                                   & 0,000000 & 177 & & I     & \textit{nlp4re\_description-EX\_AMG}                      & 0,000000 & 171 \\
    I       & \textit{nlp4re\_description-DA\_INC}                                   & 0,000000 & 178 & & I     & \textit{nlp4re\_description-EX\_ENT}                      & 0,000000 & 172 \\
    I       & \textit{nlp4re\_description-DA\_SRC}                                   & 0,000000 & 179 & & I     & \textit{nlp4re\_description-EX\_RDS}                      & 0,000000 & 173 \\
    T2T     & \textit{buggy\_similarity-avg\_similarity\_euclidean\_distance\_title} & 0,000000 & 181 & & T2T   & \textit{buggy\_similarity-max\_similarity\_jaccard\_text} & 0,000000 & 175 \\
    \cmidrule(lr){1-4} \cmidrule(lr){6-9}
\end{tabularx}
            % \caption{Bottom 10 - In progress}
        \end{minipage}
        %\label{tab:table_bottom_hive_open}
    \end{subfigure}
    \vskip\baselineskip
    % CLOSED
    \begin{subfigure}[b]{0.96\textwidth}
        \begin{minipage}[t]{\linewidth} % Use linewidth within minipage
            \centering
            \scriptsize
            \begin{tabularx}{\linewidth}{lXrrrlXrr}
    % \toprule
    \multicolumn{4}{c}{\textbf{HBASE - Closed}} & & \multicolumn{4}{c}{\textbf{HIVE - Closed}} \\
    \cmidrule(lr){1-4} \cmidrule(lr){6-9}
    % \textbf{FF} & \textbf{FN} & \textbf{Mean(IGR)} & \textbf{OR} & & \textbf{FF} & \textbf{FN} & \textbf{Mean(IGR)} & \textbf{OR} \\

    \textbf{Feature Family} & \textbf{Feature Name} & \textbf{Mean(IGR)} & \textbf{Rank}  & & \textbf{Feature Family} & \textbf{Feature Name} & \textbf{Mean(IGR)} & \textbf{Rank} \\
    \cmidrule(lr){1-4} \cmidrule(lr){6-9}
    % \midrule[\heavyrulewidth] % Optional: Use a thicker midrule for visual separation
    E\_T  & \textit{latest\_commit-churn}                                          & 0,000852 & 153 & & I\_T  & \textit{activities-work\_items\_count}                    & 0,000000 & 177 \\ % & 63 \\
    I     & \textit{nlp4re\_description-DA\_WKP}                                   & 0,000645 & 156 & & I     & \textit{nlp4re\_description-DA\_CNT}                      & 0,000000 & 178 \\ % & 64 \\
    I\_T  & \textit{nlp4re\_sentiment-IT\_SUB}                                     & 0,000617 & 157 & & I     & \textit{nlp4re\_description-DA\_SRC}                      & 0,000000 & 179 \\ % & 65 \\
    JIT   & \textit{num\_commits}                                                  & 0,000402 & 162 & & I     & \textit{nlp4re\_description-DA\_WKP}                      & 0,000000 & 180 \\ % & 66 \\
    T2T   & \textit{buggy\_similarity-avg\_similarity\_euclidean\_distance\_title} & 0,000313 & 165 & & I     & \textit{nlp4re\_description-EX\_AMG}                      & 0,000000 & 181 \\ % & 67 \\
    I\_T  & \textit{activities-work\_items\_count}                                 & 0,000000 & 182 & & I     & \textit{nlp4re\_description-EX\_ENT}                      & 0,000000 & 182 \\ % & 68 \\
    I     & \textit{nlp4re\_description-DA\_CNT}                                   & 0,000000 & 183 & & I     & \textit{nlp4re\_description-EX\_RDS}                      & 0,000000 & 183 \\ % & 69 \\
    I     & \textit{nlp4re\_description-DA\_INC}                                   & 0,000000 & 184 & & JIT   & \textit{jit-author\_date-DURATION}                        & 0,000000 & 184 \\ % & 70 \\
    I     & \textit{nlp4re\_description-DA\_SRC}                                   & 0,000000 & 185 & & JIT   & \textit{num\_commits}                                     & 0,000000 & 185 \\ % & 71 \\
    JIT   & \textit{jit-author\_date-DURATION}                                     & 0,000000 & 186 & & T2T   & \textit{buggy\_similarity-max\_similarity\_jaccard\_text} & 0,000000 & 186 \\ % & 72 \\
    \cmidrule(lr){1-4} \cmidrule(lr){6-9}
\end{tabularx}
            % \caption{Bottom 10 - Closed}
        \end{minipage}
        % \label{tab:table_bottom_hive_open}
    \end{subfigure}
\end{table}

Regarding \autoref{tab:bottom_10_igr}, we note that many features have been excluded by the analysis since not measurable in a specific proximity point. For instance, all the JIT features do not appear in Open and InProgress despite having IGR as 0 by construction. Therefore, \autoref{tab:bottom_10_igr} report the bottom features among the measurable ones
\section{Discussion}\label{sec:discussions}
\subsection{\rqone}
It is intuitive that shifting the proximity point earlier, i.e., moving it to the left, decreases the accuracy of TLP, although such early predictions remain valuable in practice due to their earlier availability. We focus to observe how much the accuracy improves as proximity to the code completion increases, and whether the TLP performance at the earliest proximity point, i.e., in Open, still surpasses that of a random baseline. 
First of all, according to the statistical tests reported in \autoref{tab:rq1_friedman_test} and by the increasing gains over the random baseline with closer proximity points, as shown in \autoref{tab:rq1_gain}, \textbf{TLP accuracy improves with proximity} with statistically significant differences observed across all seven metrics except Recall. 

Focusing the improvement of TLP over a random baseline at various proximity points (see \autoref{tab:rq1_gain}), we observe that \textbf{the gain varies across different accuracy metrics}. For example, in both HBASE and HIVE, there is no observable gain in Recall, not even in the Closed setting. A possible explanation is the highly imbalanced nature of the datasets, which are skewed toward the positive class, making random classification likely to predict positives \cite{DBLP:conf/msr/KeshavarzN22}. This explanation is supported by the observed gain on Random in Specificity, which is about 200\% in the Open proximity point. Since Specificity is analogous to Recall but pertains to the negative class, its increase highlights the difficulty of correctly identifying negative instances in imbalanced datasets.

Comparing gains over the random baseline across different proximity points, we observe that the difference between Closed and InProgress is larger than the difference between InProgress and Open, across all seven metrics and in both projects. This suggests that \textbf{from a TLP prediction accuracy perspective, the InProgress proximity point is much closer to Open than to Closed}, rather than being equidistant or closer to Closed. This finding is somewhat surprising, as we initially expected InProgress to resemble Closed more closely, given that in Open very few features are measurable. One possible explanation is that in Closed, we can leverage JIT features, thus using data about the actual code changes possibly including the bug we aim to predict.

Finally, we note that even in the Open setting, TLP performs substantially better than the random baseline across all seven metrics except Recall. According to \autoref{tab:rq1_gain}, the average gain in Open is 25\% for AUC, and Specificity shows an even more remarkable improvement of 535\% in HIVE, with notable gains in HBASE also. This suggests that \textbf{TLP seems useful even before a ticket is assigned to a developer}, and that the selected features are able to capture meaningful information about bug-inducing tickets even before most attributes become measurable. This is a promising result, indicating that TLP can be effectively performed at an early stage of the ticket lifecycle, potentially enabling bug prevention.

\subsection{\rqtwo}

According to \autoref{fig:rq2_SW_igr} and \autoref{fig:rq1_SW_accuracy}, we observe that \textbf{the predictive power of features varies significantly based on feature family, proximity point, and their interaction}. Specifically, some feature families show higher predictive power than others. For instance, the Intrinsic family shows higher predictive power than the Code family across all proximity points. However, for some families, the predictive power varies with proximity. Notably, the JIT family show (by construction) no predictive power at proximity points other than Closed, yet extremely significant at Closed. Therefore, no single feature family consistently outperforms others across all proximity points. Instead, effective prediction models should dynamically adjust feature selection strategies according to the proximity point. This finding emphasises the need to leverage JIT-related features for predictions at later lifecycle stages, while maintaining a broader set of features to ensure robust predictive performance at earlier stages.

\begin{figure}[h!]
    % \vskip\baselineskip
    \begin{subfigure}[b]{0.45\textwidth}
        \includegraphics[width=\textwidth]{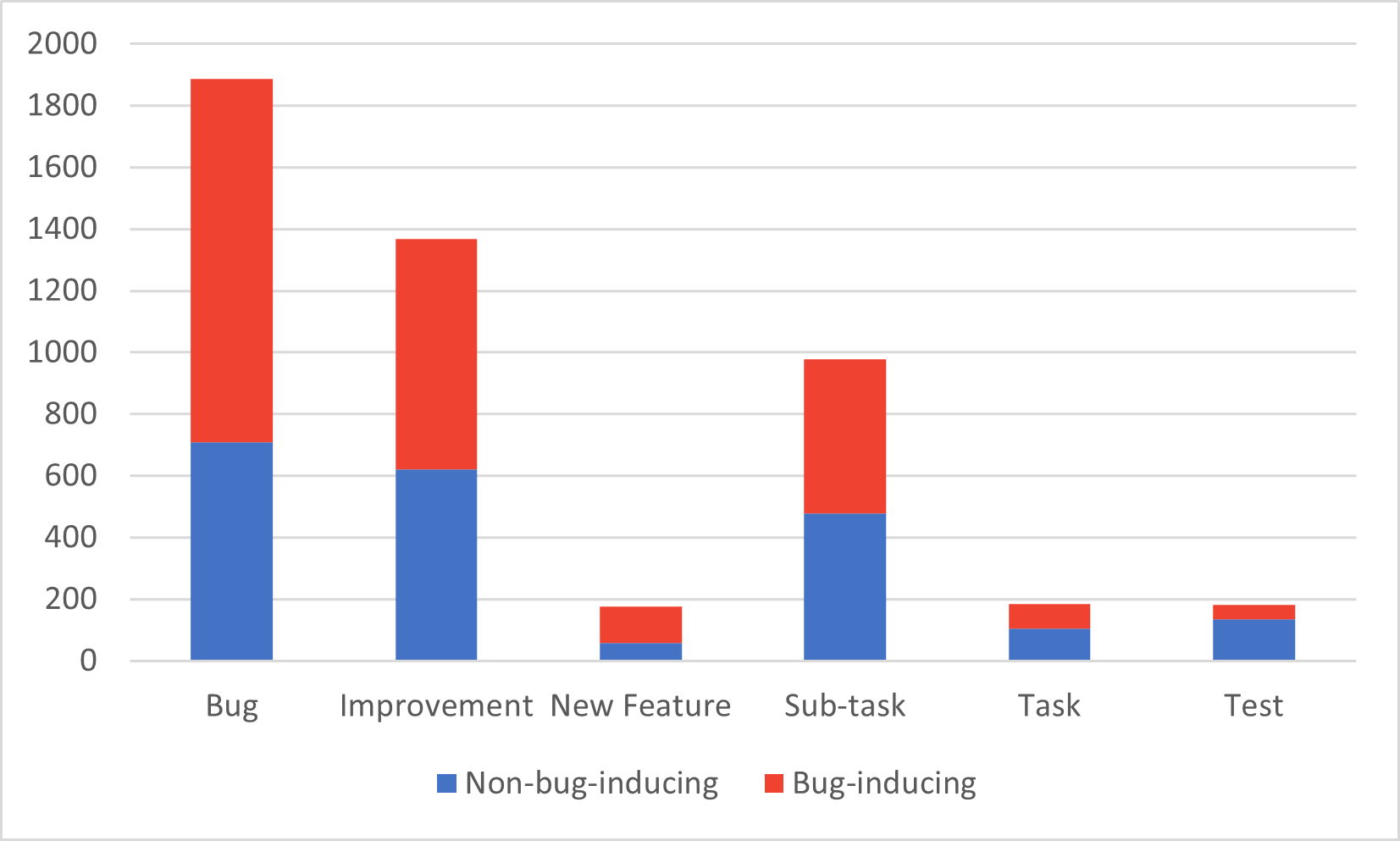}
        \caption{HBASE}
        % \label{fig:sub3}
    \end{subfigure}%
    \hfill
    \begin{subfigure}[b]{0.45\textwidth}
        \includegraphics[width=\textwidth]{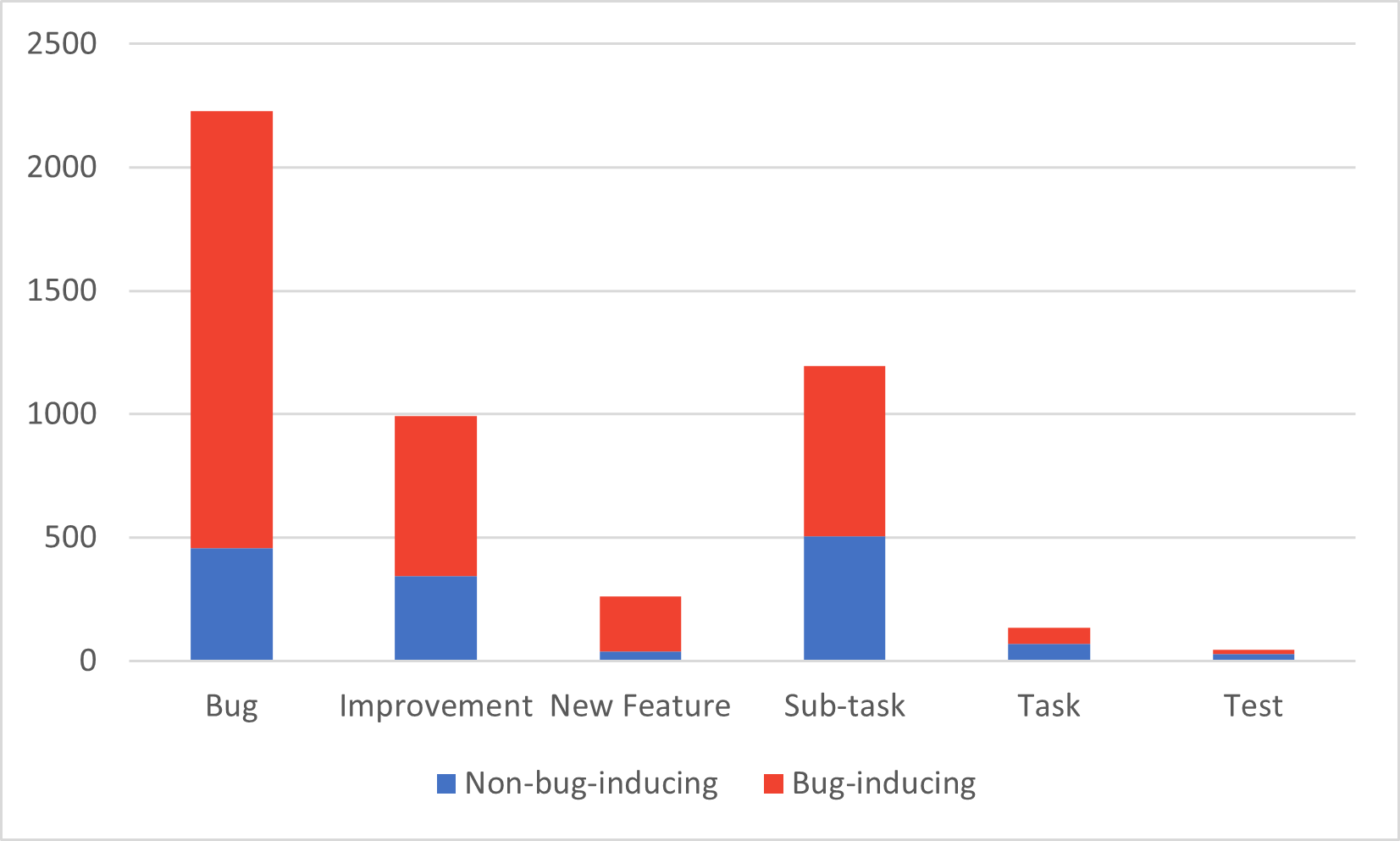}
        \caption{HIVE}
        % \label{fig:sub4}
    \end{subfigure}
    \caption{Bug-inducing issue distribution by type}
    \label{fig:bugginess-frequency-type}
\end{figure}

\begin{figure}[h!]
    \centering
    \begin{subfigure}[b]{0.45\textwidth}
        \includegraphics[width=\textwidth]{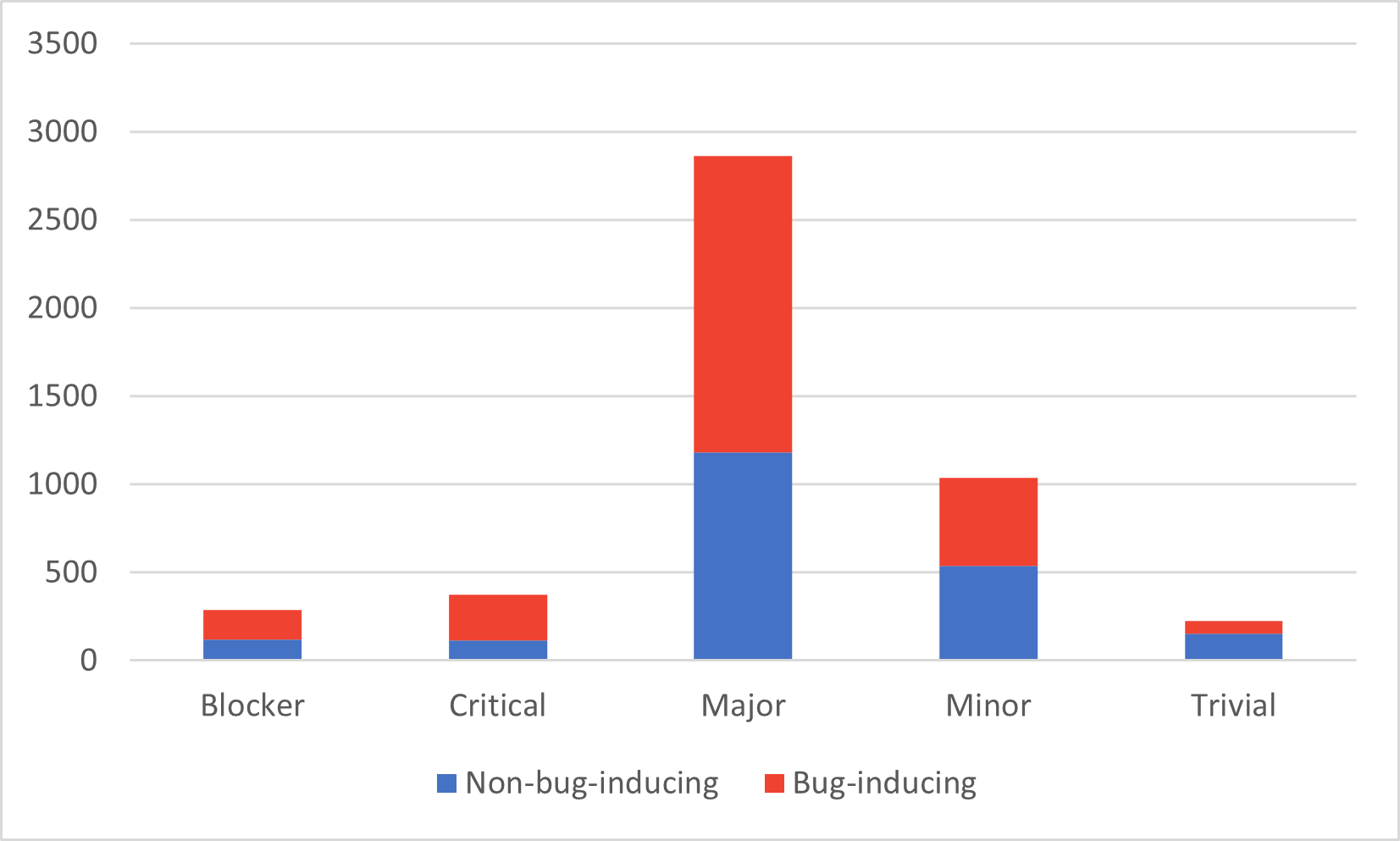}
        \caption{HBASE}
        % \caption{Distribution of issues by priority and buggy status for HBASE project}
        % \label{fig:sub1}
    \end{subfigure}%
    \hfill
    \begin{subfigure}[b]{0.45\textwidth}
        \includegraphics[width=\textwidth]{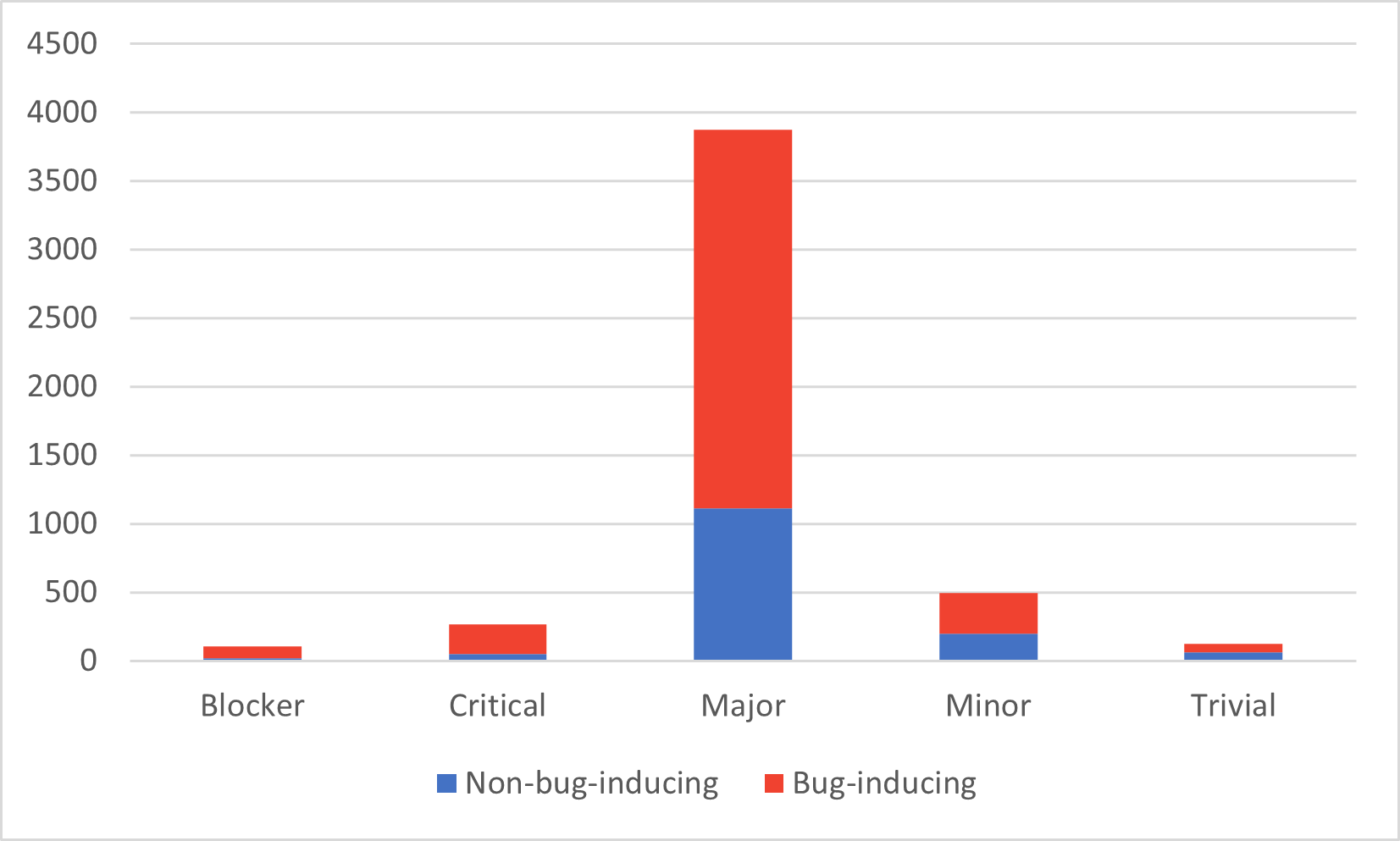}
        \caption{HIVE}
        % \caption{Distribution of issues by priority and buggy status for HIVE project}
        % \label{fig:sub2}
    \end{subfigure}
    \caption{Bug-inducing issue distribution by priority}
    \label{fig:bugginess-frequency-priority}
\end{figure}

Regarding the Top 10 features in the Open setting (\autoref{tab:bottom_10_igr}), we observe the following:

\begin{itemize} 
    \item The three most important features in both HBASE and HIVE are \textit{type}, \textit{priority}, and \textit{code\_size-number\_of\_languages}. Notably, \textit{type} exhibits an Information Gain Ratio (IGR) approximately twice as high as that of \textit{priority}, indicating its greater predictive value.
    \item With respect to \textit{type} (\autoref{fig:bugginess-frequency-type}), the absolute number of bugs is highest for tickets labeled as "bugs." However, in terms of proportion, "new features" show the highest ratio of bug-inducing tickets, suggesting that these are the most risk-prone ticket types. Conversely, "test" tickets show the lowest proportion of bug-inducing cases.

    \item For \textit{priority} (\autoref{fig:bugginess-frequency-priority}), the category associated with the most bugs in absolute numbers is "Major," while "Blocker" and "Critical" show the highest bug-inducing proportions. We can observe a general trend:  priority appears to positively correlate with proportion of bug-inducing tickets; i.e., the higher the priority the higher the proportion of bug-inducing ticket. However, it is unclear if this is due to a causation or correlation effect. A causal explanation may involve time pressure, i.e., developers may rush to complete high-priority tickets, increasing the chance of introducing bugs. Alternatively, a correlation effect may be that since high-priority tickets are more visible or widely used once implemented, they are more likely to detect and report bugs.
    
    \item Three features from the Code family also appear in the top ranks, with \textit{code\_size-total\_LOCs} being the second most important feature for both HBASE and HIVE. This suggests that the size of the codebase at the time a new ticket is opened may serve as a meaningful proxy for the likelihood of future buggy implementations.
    
    %%%% \item Regarding the NLP4RE features, there appears \textit{nlp4re\_description-DA\_ACT}, quantifying the number of distinct actions in a ticket's description. Higher values may reflect a greater implementation workload, potentially increasing the risk of bug introduction in the corresponding commit.
\end{itemize}

Regarding the Top 10 features in the InProgress and Closed settings (\autoref{tab:top_10_igr}), we observe a strong prevalence of Internal Temperature features in InProgress and JIT features in Closed. This aligns with expectations, as JIT features are only measurable in Closed. The prominence of Internal Temperature features in InProgress suggests their usefulness for TLP even before ticket closure, likely due to their ability to reflect contextual information and the potential impact of the ticket on the codebase.
Furthermore, as expected, the overall ranking of features is higher in Closed than in InProgress, indicating that features are more informative in the Closed setting. This is likely because Closed allows access to JIT features, which include data about the actual code changes—potentially including the bug-inducing modifications themselves.
Finally, it is notable that the top six features in the Closed setting are the same for both projects. This is an important finding given that the JIT family consists of 15 features, and that features from other families—such as \textit{activities-count}—exhibit higher IGR scores than some JIT features. This consistency suggests that these top-ranked features are particularly valuable for TLP and may serve as reliable predictors in a JIT-based approach even if not of the JIT family.

Regarding the features with the lowest predictive power, \autoref{tab:bottom_10_igr} shows that the JIT features \textit{num\_commits} and \textit{author\_date-duration} rank among the lowest, even in the Closed setting where JIT features are fully available. This suggests the need for feature selection within the JIT family and highlights the potential value of incorporating TLP-specific features to enhance prediction in JIT-based models.

\section{Related Work}\label{sec:related}
This section positions our work within six key research areas: requirements quality, developer-related metrics, change impact analysis, defect prediction, temporal proximity in prediction, and natural language processing (NLP) applied to software tickets.

\subsection{Requirements Quality}
Requirements quality has long been identified as a determinant of downstream software quality. Past works define high-quality requirements as unambiguous, complete, consistent, and testable \cite{berry1998requirements, DBLP:conf/icse/WilsonRH97, wiegers2013software}. Deficiencies in these characteristics, i.e., requirements smells \cite{DBLP:conf/profes/GentiliF23}, can propagate into implementation, and impact defects and maintenance costs \cite{DBLP:journals/jss/AhonenS10, boehm2007software, DBLP:conf/re/KamataT07}.
Past works proposed two strategies for mitigating these smells. The first strategy is called  preventive, i.e, the promoting high-quality authoring through structured natural language standards (e.g., INCOSE \cite{incose2023incose}, ISO/IEC/IEEE 29148 \cite{iso_iec_ieee_29148}). The second strategy is corrective, i.e., using automated smell detectors based on NLP techniques \cite{DBLP:journals/jss/FemmerFWE17, ferrari2018detecting}. Both strategies assume the existence of formally specified requirements, which is a condition rarely met in open-source development, where issue tracking systems (ITSs) like Jira are used  \cite{li2018jira}. Tickets of open-source projects usually mix functional intent and implementation detail with highly variable structure.
Given this lack of standardization, our work diverges from rule-based smell detection and instead models the textual and structural properties of tickets themselves, including features like syntactic completeness, ambiguity, and task granularity. This follows earlier work on more general quality proxies for informal specifications \cite{DBLP:journals/isse/CarlsonL14, DBLP:conf/icse/WilsonRH97}. While these broader features are less actionable than fine-grained smell types, they are more appropriate for predicting defect-inducing changes in non-standardized settings.

\subsection{Developer-Centric Metrics}
Developers play a central role in defect introduction. Prior work has shown that factors such as insufficient code ownership, poor responsibility assignment, and variability in developer expertise are strongly associated with fault-proneness \cite{DBLP:journals/tse/BergersenSD14, DBLP:journals/tse/LeeNHKI16}. In particular, \citep{DBLP:conf/promise/MatsumotoKMMN10} demonstrated the predictive utility of developer-level metrics such as commit frequency and file ownership.
Our work extends this line by including both individual and collective developer-related features, such as the number of developers assigned to a module, author diversity in ticket discussions, and historical defect rates of individual developers. These metrics aim to capture socio-technical dimensions of software quality in the context of ticket-level analysis.

\subsection{Change Impact Analysis}
Change impact analysis (CIA) aims to predict the effects of modifications in software systems, hence supporting better planning and quality assurance \cite{DBLP:conf/iwpse/Lehnert11, DBLP:conf/sbes/BordinB18, DBLP:journals/access/AnwerWWM19}. Past studies have introduced both taxonomies and automated methods for traceability and impact estimation, mainly focusing on predicting code-level changes \cite{DBLP:conf/iwpc/AungHS20, DBLP:conf/icsoft/GentiliCF24}.
In this work we adopt a ticket-centric view of CIA by predicting the effect of implementing a ticket in terms of inducing a bug. 

\subsection{Defect Prediction}
Defect prediction remains a central research area in empirical software engineering. A large body of work has focused on Just-In-Time Defect Prediction (JITDP), where models assess the likelihood that a given code change will introduce a defect \cite{DBLP:conf/wcre/KameiS16, DBLP:journals/csur/ZhaoDC23}. These models rely heavily on change-level metrics and their performance is influenced by data quality \cite{DBLP:journals/ase/LiDZJW24}, parameter tuning \cite{DBLP:journals/infsof/FuMS16}, and temporal retraining frequency \cite{DBLP:journals/tosem/FalessiAP22, DBLP:journals/tse/McIntoshK18}.
Our work takes a different angle by shifting the granularity of prediction from commits to tickets. While prior studies have shown the predictive value of requirements and design artefact quality for early-stage defect prediction \cite{DBLP:journals/jss/OzakinciT18, DBLP:journals/ese/FalessiLCEC23}, few have formalized this at the level of ITS tickets. We construct a ticket-level dataset derived from manually validated JIT data and integrate static code, developer, and textual features into a unified prediction framework. Our study also explicitly models prediction performance at multiple lifecycle stages (Open, In Progress, Closed), extending the evaluation protocol used in recent benchmark studies (e.g., \cite{DBLP:journals/corr/abs-1809-01510}).

\subsection{Temporal Proximity in Prediction}
The timing of prediction affects its accuracy, as widely demonstrated in fields ranging from weather forecasting to financial modelling \cite{box2015time, orrell2001model, brockwell2002introduction}. Empirical and theoretical studies confirm that predictive performance improves with decreasing temporal distance to the target event due to better information availability and reduced stochastic uncertainty \cite{lorenz1963deterministic, shumway2017arima, graves2012long}.
Our work is the first to operationalize this principle in the context of TLP. We define and compare model performance across three ticket lifecycle stages (creation, assignment, and closure), thus quantifying the value of temporal proximity in practical software defect prediction. This design choice is methodologically aligned with temporal weighting approaches such as ARIMA and LSTM, prioritising recent observations in time-series prediction \cite{florita2009comparison, grover2015deep}.

\subsection{NLP-Based Analysis of Tickets}
The widespread use of ITSs such as Jira led to an explosion of unstructured ticket data. To manage this complexity, previous studies applied natural language processing to support ticket classification, prioritization, severity estimation, and duplicate detection \cite{DBLP:journals/tse/SunGYLLZ24, DBLP:conf/cascon/AntoniolAPKG08a, DBLP:journals/access/AhmedBS21, DBLP:journals/jss/ZhangCYLL16, DBLP:journals/tse/ChoetkiertikulD19, DBLP:journals/tse/FalessiCC13}. Building on this foundation, recent work has shown that the textual similarity between new and historical tickets can serve as a proxy for defect risk, particularly when linked to specific code components \cite{DBLP:journals/tse/FalessiRGC20}. Inspired by this, we treat feature request tickets as informal requirements and apply NLP methods to extract syntactic and semantic features. These include both shallow linguistic metrics and semantic similarity measures, which are incorporated into our predictive models. Further details are provided in \autoref{sec:features}.

\section{Threats to Validity}\label{sec:threats}
Following established guidelines for empirical research in software engineering \cite{DBLP:books/sp/WohlinRHORW24}, we discuss the threats to validity in four categories: conclusion, internal, construct, and external validity.

\subsection{Conclusion Validity}

Conclusion validity concerns the degree to which conclusions regarding the relationships between the treatment and the outcome are statistically sound and justified \cite{DBLP:books/sp/WohlinRHORW24}. In this study, we used the non-parametric Friedman test \cite{Friedman1937} to evaluate our hypotheses. Non-parametric tests are less sensitive to assumptions about data distribution and, hence, more appropriate given the nature of defect prediction datasets.While non-parametric tests are typically more conservative and may increase the risk of Type II errors, i.e., failing to reject a false null hypothesis, our results consistently reject the null, suggesting robust differences. We deliberately avoided parametric alternatives such as ANOVA, which assume normality and homoscedasticity and are more susceptible to Type I errors.

\subsection{Internal Validity}

Internal validity refers to the extent to which observed effects can be causally attributed to the treatments rather than confounding variables \cite{DBLP:books/sp/WohlinRHORW24}. A primary threat arises from the absence of ground truth for ticket defectiveness, which could bias the labelling of defect-inducing tickets. To mitigate this, we relied on projects previously vetted in JIT defect prediction studies, e.g., \cite{DBLP:journals/tse/KameiSAHMSU13, DBLP:journals/ese/PatelAH24}, where established methodologies for linking commits to issue trackers were applied. Furthermore, we did not differentiate among severity levels of the defects, in line with common practice in defect prediction research, e.g., \cite{DBLP:journals/tse/Tantithamthavorn19, DBLP:journals/ese/FalessiLCEC23}. While this choice may hide finer-grained distinctions in bug impact, it prevents potential confounding due to inconsistent severity labelling and it helps preserve the internal consistency of the bug label definition across projects.

\subsection{Construct Validity}

Construct validity addresses the degree to which the operational measures used in the study accurately capture the intended theoretical constructs \cite{DBLP:books/sp/WohlinRHORW24}. This threat is particularly relevant in TLP due to the reliance on derived metrics and preprocessed features. To minimize this threat, we based our feature engineering and evaluation design on prior work validated in recent JIT literature \cite{DBLP:journals/ese/PatelAH24}. One specific concern relates to the decision to avoid hyperparameter tuning of the classifiers. While hyperparameter optimization is known to improve predictive performance \cite{DBLP:journals/infsof/FuMS16}, our goal was not to maximize performance per se, but rather to compare the relative efficacy of TLP across proximity points and feature families. Consequently, using default hyperparameters ensures experimental consistency and avoids confounding due to model overfitting, particularly in a temporally ordered evaluation setting.

\subsection{External Validity}

External validity concerns how the results can be generalized beyond the studied context  \cite{DBLP:books/sp/WohlinRHORW24}. Our study includes two open-source systems with a high commit-to-ticket linkage rate, ensuring high-quality ground truth but limiting the generalizability of the findings to other contexts. The choice to restrict the dataset was intentional: to maintain internal validity and control over data quality. Nonetheless, additional replications across diverse systems—particularly those with different domain characteristics, team structures, or development methodologies—are needed to assess the generalizability of the observed results.  Additionally, given the rapid emergence of generative AI for software development \cite{takerngsaksiri2025humanintheloopsoftwaredevelopmentagents, DBLP:journals/tosem/HouZLYWLLLGW24, DBLP:journals/software/CarletonFZX24}, future research should investigate customized TLP approaches for automatically generated code. To foster reproducibility and enable external validation, all datasets, scripts, and results are made publicly available\footnote{\url{https://doi.org/10.5281/zenodo.15341225}}.

\section{Conclusion and Future Work}\label{sec:conclusion}
Following the principle that prevention is preferable to remediation, this work introduces and evaluates the first approach to Ticket-Level (bug) Prediction, i.e., a method designed to identify tickets that are likely to introduce bugs upon implementation. The study examines how prediction accuracy and the discriminative power of features evolve in relation to temporal proximity within the ticket lifecycle. Specifically, we investigate three temporal points corresponding to distinct stages in the lifecycle of a ticket: Open, In Progress, and Closed. The central premise underpinning this effort is that earlier predictions, even if less accurate, may have higher utility due to their greater potential for enabling proactive quality assurance. Our TLP approach leverages 72 features belonging to six different families: code, developer, external temperature, internal temperature, intrinsic, ticket to tickets, and JIT.

Our first major finding is that TLP accuracy improves with temporal proximity to ticket closure, with statistically significant differences across most evaluation metrics. However, even at the earliest stage, i.e., Open, when very limited information is available or accurate, TLP models consistently outperform a random baseline in most metrics, especially in AUC and Specificity. This result demonstrates that meaningful predictive info can be extracted from ticket metadata and static attributes at ticket creation time. This result complements and extends the literature on bug prediction at class, method or commit level \cite{DBLP:conf/wcre/KameiS16} by showing that defect prediction can be effectively moved upstream, offering opportunities for risk-aware ticket triaging and developer assignment before any code is written.
Unexpectedly, we observed that In Progress is closer to Open than Closed in terms of predictive accuracy. This finding challenges intuitive assumptions and suggests that the transition from Open to In Progress provides only marginal gains in predictive signal—likely due to the limited availability of new features until the ticket is closed. In contrast, the Closed stage benefits from JIT features derived from actual code changes, including those potentially introducing the bugs themselves. From a practical standpoint, this implies that the major leap in prediction capability occurs only upon ticket closure, and that pre-closure models should focus on enhancing signal quality without code-level features.
A key contribution of this study is showing that no single feature family is uniformly dominant across all stages. Instead, the predictive utility of features is highly contextual: Intrinsic and Code features are most useful early in the lifecycle, Internal Temperature features gain prominence in In Progress, and JIT features dominate in the Closed stage. These findings align with and extend prior work advocating for adaptive, project-sensitive defect prediction models \cite{DBLP:journals/tse/Tantithamthavorn/2016}. They highlight the need for lifecycle-aware prediction systems that adjust their feature sets dynamically as more information becomes available. Moreover, we show that not all JIT features are equally informative; several rank consistently low even when fully available. This underscores the need for feature selection within strong feature families, a point also raised in earlier studies of defect model stability \cite{DBLP:conf/icse/GhotraMH15}.
Our feature-level analysis further reveals that ticket type and priority are among the most important predictors at early stages, with “new feature” tickets showing the highest proportion of bug-inducing outcomes. Higher-priority tickets (e.g., “Blocker”, “Critical”) also exhibit elevated bug rates, possibly due to time pressure or heightened visibility. These findings support and refine prior observations regarding the relationship between process factors and defect risk \cite{DBLP:conf/sigsoft/ZimmermannNGGM09, DBLP:conf/icse/HerzigJZ13}. From a practical perspective, these attributes offer immediate utility in early-stage triaging, allowing teams to prioritize review and testing resources based on easily available ticket descriptors.
An additional noteworthy observation is the cross-project consistency of top-performing JIT features. Despite differences in project characteristics, the top-ranked JIT features at the Closed stage are nearly identical for both HBASE and HIVE, suggesting the existence of robust, transferable signals within this family. However, this consistency contrasts with the observed variability among low-ranking JIT features, indicating that future JIT models may benefit from ticket features.

Regarding future work, this study opens several avenues for advancing Ticket-Level Bug Prediction along methodological and practical dimensions. First, we aim to improve the external validity of TLP by extending our analysis to a broader set of open-source and industrial projects. This will enable investigation into how project characteristics—such as team structure and development process—affect model performance and feature behaviour.
Second, we will pursue refinements in feature engineering, exploring new families such as developer activity metrics and repository evolution indicators. We also plan to leverage deep learning techniques, including domain-specific language models, to extract semantically rich features from textual and code artefacts.
Third, we will evaluate ensemble methods and hybrid architectures that combine traditional machine learning with deep models to enhance predictive performance. Real-time prediction capabilities will be explored to enable integration with continuous development pipelines.
Fourth, we will investigate the impact of concept drift on TLP stability and develop adaptive learning strategies to maintain accuracy over time, including dynamic retraining and sliding window approaches.
Finally, we envision applying TLP in practice by designing developer-facing tools and embedding predictions into issue tracking and CI systems. We also plan to explore its use in software education, providing real-time feedback to students on risk-prone development patterns.

\printbibliography
\appendix
\section{Appendix}
\label{sec:Appendix}

\subsection{Feature families}
\label{sec:Featurefamilies}
%\begin{minipage}[t]{\linewidth}
\subsubsection{Code-Level Features}
% Code-related factors capture intrinsic characteristics of the underlying system that influence the ease and risk of implementing a ticket. This includes static quality metrics and structural attributes that correlate with maintainability and defect-proneness.

% \paragraph{Code Quality.}
\begin{itemize}
    \item \textbf{Smell Count} (\textit{code\_quality-smells\_count}): Total number of PMD rule violations in the codebase at the time of ticket creation.

    \item \textbf{Total LOCs} (\textit{code\_size-total\_LOCs}): Total number of lines of code.
    \item \textbf{Number of Files} (\textit{code\_size-number\_of\_files}): Total number of source files.
    \item \textbf{Number of Languages} (\textit{code\_size-number\_of\_languages}): Number of programming languages used in the codebase.
\end{itemize}

\subsubsection{Developer-Related Features}
\begin{itemize}
    \item \textbf{ANFIC} (\textit{assignee-ANFIC}): Ratio of bug-inducing tickets to total tickets assigned to the developer \cite{DBLP:conf/promise/MatsumotoKMMN10}.
    \item \textbf{Familiarity} (\textit{assignee-familiarity}): Proportion of tickets assigned to the developer relative to the total number in the project.
\end{itemize}

\subsubsection{External Temperature}
\begin{itemize}
    \item \textbf{Temporal Locality} (\textit{temporal\_locality}): Ratio of recent bug-inducing tickets in a fixed temporal window.
    \item \textbf{Weighted Temporal Locality} (\textit{temporal\_locality-weighted}): Weighted ratio, giving higher weight to temporally closer bugs.
    \item \textbf{Commits During Progress} (\textit{commits\_while\_in\_progress-count}): Number of commits submitted while the ticket was in progress.
    \item \textbf{Churn During Progress} (\textit{commits\_while\_in\_progress-churn}): Cumulative LOCs added, modified, or deleted during the same period.
    \item \textbf{Latest Commit Churn} (\textit{latest\_commit-churn}) and \textbf{File Count} (\textit{latest\_commit-number\_of\_files}): Change size and scope of the latest commit preceding the ticket.
\end{itemize}

\subsubsection{Internal Temperature}
\begin{itemize}
    \item \textbf{Ticket Participants Count} (\textit{issue\_participants-count}): The more the developers working on a software module, the higher the chance a defect is injected as a result \citep{DBLP:conf/sigsoft/PinzgerNM08}. We measure the number of participants involved in the ticket implementation, i.e. authors of changelog entries, reporter, assignee, creator.

    \item \textbf{Activities Count} (\textit{activities-count}): Developers tend to discuss problematic software entities more \citep{DBLP:conf/fase/BacchelliDL10}. We transfer this concept to the ticket level by counting ticket Activities. Activities in a ticket can be comments, work items and histories.

    \item \textbf{Comments Count} (\textit{activities-comments}): Ticket participants use comments to express their opinions, ask for clarifications, provide additional information, etc.

    \item \textbf{Work Items Count} (\textit{activities-work\_items\_count}): Work items are used to track the time spent by participants on the ticket.

    \item \textbf{Histories Count} (\textit{activities-histories}): Histories are used to track the changes made to the ticket.

    \item \textbf{Sentiment Polarity}              (\textit{nlp4re\_sentiment-IT\_POL}): measures the positiveness (or negativeness) of a sentence. It is a number lying between -1 (extremely negative) and 1(extremely positive) \cite{DBLP:reference/ml/0016017}.

    \item \textbf{Sentiment Subjectivity}          (\textit{nlp4re\_sentiment-IT\_SUB}): measures the amount of personal opinion and feelings with respect to factual information contained in a sentence. Typically, the higher the index, the less objective is the language used in a sentence \cite{DBLP:reference/ml/0016017}. It is a number lying between $[0;1]$.

    \item \textbf{Number Of Negative Comments}        (\textit{nlp4re\_sentiment-CM\_NNS}): measures the occurrence of negative comments associated to a requirement

    \item \textbf{Percentage Of Negative Comments} (\textit{nlp4re\_sentiment-CM\_PNS}): measures the occurrence of negative comments divided by the total number fo comments 

    \item \textbf{Presence Of One Negative Comment}(\textit{nlp4re\_sentiment-CM\_ONS}): measures the presence of at least one negative comment

\end{itemize}

\subsubsection{Intrinsic}

% Intuitively, some tickets can be considered inherently more difficult to implement than others.
\begin{itemize} 
    \item  \textbf{Priority} (\textit{priority}): We measure the priority of the ticket, namely a level of importance telling what ticket should be implemented first. Prioritizing one ticket over another means allocating more time to the former at the cost of the latter, hence the latter could be subject to a rushed development which could produce a buggy implementation. Besides, the higher the priority, the more urgent the ticket is, the more the stress can burden the assignee, leading to a higher chance of mistakes.

    \item  \textbf{Components Count} (\textit{components-count}): This feature is inspired by the Entropy feature described in \citet{DBLP:journals/ese/PatelAH24}. We count the number of project components the ticket is related to, in order to measure the dispersion of the required changes.

    \item  \textbf{Components Max Bugginess}(\textit{components-max\_bugginess}): We measure the highest bugginess among the components the ticket is related to. We compute the bugginess of the component by counting the number of bug-inducing ticket historically related to the component divided by the total number of tickets related to the component until the measurement date. The idea is that a historically buggy component could be inherently fragile, hence further changes to it could induce more bugs. 

    \item  \textbf{Type}(\textit{type}): It has been shown that often bug fixing changes induce more bugs in the code \citep{DBLP:conf/icse/GuBHS10}. We extend this concept by considering the change type of the ticket, i.e: bug, improvement, new feature, subtask, etc. It is worth noting than the activity of finding and fixing bugs, although is the most rewarding when successful, can be very frustrating and stressful \citep{DBLP:journals/tse/WinterBCHHNW23}, leading to a higher chance of mistakes.

    \item  \textbf{Description Attribute Action}       \textit{(nlp4re\_description-DA\_ACT)}: measures the occurrence of actions by applying patterns to detect obligations and compound verb phrases \cite{incose2023incose}. The higher, the more probable is that the current requirement should be split into multiple requirements. 

    \item  \textbf{Description Attribute Conditionals} \textit{(nlp4re\_description-DA\_CND)}: measures the occurrence of conditional patterns in a sentence, such as the presence of words like "if," "when," "unless," and "depends on," among others \cite{DBLP:conf/icse/WilsonRH97, DBLP:conf/issre/JiangCM07}

    \item  \textbf{Description Attribute Continuances} \textit{(nlp4re\_description-DA\_CNT)}: measures the occurrence of continuance indicators in a sentence, including phrases like "see below," "as follows," "listed," and "in particular," among others \cite{DBLP:conf/icse/WilsonRH97, DBLP:conf/issre/JiangCM07}

    \item  \textbf{Description Attribute Imperatives}  \textit{(nlp4re\_description-DA\_IMP)}: measures the occurrence of imperative expressions in a sentence, such as "shall," "must," and "is required to," among others \cite{DBLP:conf/icse/WilsonRH97, DBLP:conf/issre/JiangCM07}

    \item  \textbf{Description Attribute Incompletes}  \textit{(nlp4re\_description-DA\_INC)}: measures the occurrence of incomplete markers, identifying acronyms like "TBD," "TBR," "TBC," and "TODO," which indicate missing or pending content \cite{DBLP:conf/icse/WilsonRH97, DBLP:conf/issre/JiangCM07}

    \item  \textbf{Description Attribute Options}      \textit{(nlp4re\_description-DA\_OPT)}: measures the occurrence of the use of optionality markers, including words like "can," "could," "may," and "optionally," which indicate non-mandatory elements \cite{DBLP:conf/icse/WilsonRH97, DBLP:conf/issre/JiangCM07}

    \item  \textbf{Description Attribute Sources}      \textit{(nlp4re\_description-DA\_SRC)}: measures the occurrence of references to external resources, like files and websites \cite{incose2023incose}

    \item \textbf{Description Attribute Weak Phrases} \textit{(nlp4re\_description-DA\_WKP)}: measures the occurrence of vague or non-assertive phrases that may weaken the clarity and precision of a sentence, e.g. "adequate", "as a minimum", "be capable of" and so on \cite{DBLP:conf/icse/WilsonRH97, DBLP:conf/issre/JiangCM07}

    \item \textbf{Description Attribute Risk Level}   \textit{(nlp4re\_description-DA\_RKL)}: measures the overall level of risk associated to each requirement by summing up each previous index "DA\_<>"

    \item \textbf{Number Of Subjects}                 \textit{(nlp4re\_description-EX\_SBJ)}: measures the number of general nouns in sentence, identifying subjects and objects \cite{incose2023incose}

    \item \textbf{Number Of Words}                    \textit{(nlp4re\_description-EX\_CNS)}: measures the number of words in sentence \cite{DBLP:conf/icse/WilsonRH97}

    \item \textbf{Number Of Verbs}                    \textit{(nlp4re\_description-EX\_VRB)}: measures the occurrence of verbs \cite{incose2023incose}

    \item \textbf{Number Of Ambiguities}               \textit{(nlp4re\_description-EX\_AMG)}: measures the number of ambiguos words used in the sentences, e.g. "some", "many", "few", "often" and so on. \cite{incose2023incose}

    \item \textbf{Number Of Directives}               \textit{(nlp4re\_description-EX\_DIR)}: measures the use of directives markers that represent instructions or references, such as "e.g.," "i.e.," "figure," "table," "for example," and "note." \cite{DBLP:conf/icse/WilsonRH97}

    \item \textbf{Readability Score}                  \textit{(nlp4re\_description-EX\_RDS)}: measures how readable a piece of text by applying the "Flesch reading ease" score. The lowest is the score, the more technical is the language used in the sentence. \cite{DBLP:conf/icse/WilsonRH97}

    \item \textbf{Sentence Completeness}              \textit{(nlp4re\_description-EX\_ICP)}: measures whether a sentence is syntactically complete based on the presence of a nominal subject, a verb, and an object (direct, indirect, or prepositional) \cite{DBLP:conf/icse/WilsonRH97}

    \item \textbf{Action Density}                     \textit{(nlp4re\_description-EX\_ACD)}: measures the number of actions with respect the total number of words \cite{incose2023incose} 

% https://www.clarin.eu/resource-families/tools-named-entity-recognition
\item \textbf{Number Of Entities}                 \textit{(nlp4re\_description-EX\_ENT)}: measures the number of recognized named entity (NER) in a sentence \cite{incose2023incose}
\end{itemize}

\subsubsection{Ticket-to-Ticket Similarity (T2T)}
\begin{itemize}
    \item \textbf{Max Cosine Similarity (Title)}: Maximum cosine similarity between the TF-IDF vector of the current ticket title and those of past bug-inducing ticket titles.
    \item \textbf{Max Cosine Similarity (Description)}: Maximum cosine similarity between ticket descriptions.
    \item \textbf{Avg Cosine Similarity (Title)}: Average cosine similarity between the current ticket title and past buggy titles.
    \item \textbf{Avg Cosine Similarity (Description)}: Average cosine similarity over descriptions.
    \item \textbf{Max Jaccard Similarity (Title)}: Maximum Jaccard similarity between token sets of ticket titles.
    \item \textbf{Max Jaccard Similarity (Description)}: Maximum Jaccard similarity between token sets of descriptions.
    \item \textbf{Avg Jaccard Similarity (Title)}: Average Jaccard similarity across titles.
    \item \textbf{Avg Jaccard Similarity (Description)}: Average Jaccard similarity across descriptions.
    \item \textbf{Max Euclidean Distance (Title)}: Maximum Euclidean distance between term frequency vectors of ticket titles.
    \item \textbf{Max Euclidean Distance (Description)}: Maximum distance between descriptions.
    \item \textbf{Avg Euclidean Distance (Title)}: Mean Euclidean distance across all title comparisons.
    \item \textbf{Avg Euclidean Distance (Description)}: Mean distance over all description comparisons.
\end{itemize}

\subsubsection{JIT}
%When describing a feaure, remember to report the aggregation method.  "In case of multiple commit we aggregated the value by using xxx."
\begin{itemize}
    \item \textbf{Authors Count} (\textit{jit-ndev-MAX}): We count the highest number of authors involved in a commit linked to the ticket.

    \item \textbf{Developer Recent Experience} (\textit{jit-arexp-MIN}): We measure the lowest recent experience of the authors involved in the commits linked to the ticket. 

    \item \textbf{Developer Experience} (\textit{jit-aexp-MIN}): We measure the lowest experience of the authors involved in the commits linked to the ticket.

    \item \textbf{Developer Subsystem Experience} (\textit{jit-asexp-MIN}): We measure the lowest subsystem experience of the authors involved in the commits linked to the ticket.

    \item \textbf{Modified Subsystems Count} (\textit{jit-ns-MAX}): We count the highest number of subsystems modified by a commit linked to the ticket.

    \item \textbf{Age} (\textit{jit-age-MIN}): We measure the lowest temporal distance between a commit linked to the ticket and the most recent commit preceding it.

    \item \textbf{Author Date} (\textit{jit-author\_date-DURATION}): We measure the time span between the earliest and the latest commit linked to the ticket.

    \item \textbf{LOCs Added} (\textit{jit-la-SUM}): We sum the LOCs added by all commits linked to the ticket.

    \item \textbf{LOCs Deleted} (\textit{jit-ld-SUM}): We sum the LOCs deleted by all commits linked to the ticket.

    \item \textbf{Type} (\textit{jit-fix-COUNT\_TRUE}): We count how many fixing commits are linked to the ticket. Fixing changes are more prone to introduce bugs \citep{DBLP:conf/icse/GuBHS10}.

    \item \textbf{Modified Directories Count} (\textit{jit-nd-MAX}): We count the highest number of directories modified by a commit linked to the ticket.

    \item \textbf{Unique Changes Count} (\textit{jit-nuc-MAX}): We count the highest number of unique changes made by a commit linked to the ticket.

    \item \textbf{Entropy} (\textit{jit-ent-MAX}): We measure the highest entropy of a commit linked to the ticket.

    \item \textbf{Modified Files Count} (\textit{jit-nf-MAX}): We count the highest number of files modified by a commit linked to the ticket.

    \item \textbf{Number of Commits} (\textit{num\_commits}): we count the number of commits linked to the ticket to distinguish the cases when commits with different values for every JIT feature produce same values when aggregated.
\end{itemize}

\newpage
\subsection{Setups Comparison}
\label{sec:SetupsComparison}

\begin{figure}[h]
    \centering
    \includegraphics[width=\linewidth]{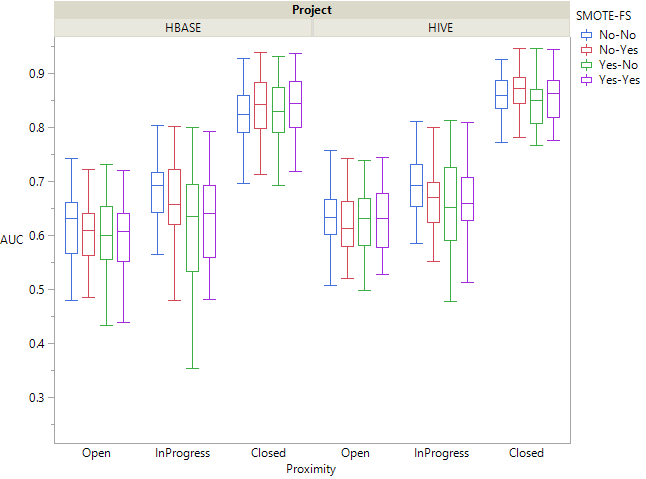}
    \caption{Average AUC accuracy: moving window approach with and without feature selection and with and without SMOTE balancing.}
    \label{fig:setup_comparison_mw2}
\end{figure}

Figure \ref{fig:setup_comparison_mw2} reports the average AUC accuracy for the different setups, namely the moving window approach with and without feature selection and with and without SMOTE for balancing.

\end{document}